\documentclass[a4paper,fleqn,usenatbib]{mnras}

\usepackage{newtxtext,newtxmath}
\usepackage{amsmath}

\usepackage[T1]{fontenc}
\usepackage{ae,aecompl}

\usepackage{graphicx}	
\usepackage{amsmath}	
\usepackage{amssymb}	
\usepackage{color,soul}

\newcommand{\kms}{\,km\,s$^{-1}$}
\newcommand{\co}[2]{$^{#1#2}$CO}
\newcommand{\myr}{\,M$_{\sun}$\,yr$^{-1}$}

\title[Molecular gas in MGE 042.0787+00.5084]{A slowly expanding torus associated with the candidate LBV MGE 042.0787+00.5084}

\author[C. Bordiu et al.]{
Cristobal Bordiu,$^{1}$\thanks{E-mail: cbordiu@cab.inta-csic.es}
J. Ricardo Rizzo,$^{1}$
Alessia Ritacco$^{2}$
\\
$^{1}$Centro de Astrobiolog\'ia (INTA-CSIC), Ctra. M-108, km. 4, 28850 Torrej\'on de Ardoz, Madrid, Spain\\
$^{2}$Institut de RadioAstronomie Millim\'etrique (IRAM), Granada, Spain
}

\date{Accepted XXX. Received YYY; in original form ZZZ}

\pubyear{2018}
\begin{document}
\label{firstpage}
\pagerange{\pageref{firstpage}--\pageref{lastpage}}
\maketitle

\begin{abstract}
The luminous blue variable (LBV) phase is a poorly understood stage in the evolution of high mass stars, characterized for its brevity and instability. The surroundings of LBV stars are excellent test beds to study their dense stellar winds and eruptive mass-loss events. Aiming to  improve our knowledge of the LBV phase, we observed the $J=1\rightarrow0$ and $J=2\rightarrow1$ lines of CO and \co13 in a field of $1\farcm5 \times1\farcm5$ around the recently identified candidate LBV MGE 042.0787+00.5084, using the IRAM 30-m radio telescope. We report the first detection of molecular emission associated with this source, tracing a structure with an evident circumstellar distribution. Morphology and kinematics of the gas can be explained by an expanding torus, a structure that may have originated from stellar ejecta or the action of stellar winds onto the parent molecular cloud. We derive the physical properties of the gas by means of LTE and non-LTE line modelling, obtaining densities of H$_2$ in the order of $10^3$ cm$^{-3}$ and kinetic temperatures below 100 K. In addition, we build a kinematic model to reproduce the structure and velocity fields of the gas, which is in good agreement with the observations. We estimate a total molecular gas mass of $0.6 \pm 0.1$ M$_{\sun}$ and a dynamical age of $6\times10^4$ years, leading to an average mass-loss rate of 0.8--1.2$\times10^{-5}$\myr.

\end{abstract}

\begin{keywords}
stars: mass-loss -- stars: massive -- stars: evolution -- ISM: molecules -- ISM: kinematics and dynamics -- ISM: clouds
\end{keywords}



\section{Introduction} \label{sec:introduction}

High mass stars strongly disturb the composition and structure of the interstellar medium by delivering large amounts of thermal and mechanical energy into their surroundings.

Before exploding as core-collapse supernovae, high mass stars evolve through a series of short-lived and violent phases \citep{Langer1994}. Among these transitional stages, the so-called luminous blue variable phase (hereafter LBV) is possibly the most puzzling due to its intrinsic instability. LBV stars are luminous and hot supergiants characterized by dense winds that sustain high mass loss rates --sometimes reaching $10^{-3}$ \myr \citep{VanBoekel2003}--. This steady mass loss is occasionally accompanied by violent outbursts that eject the outer stellar layers into the ISM, like the eruptions of $\eta$ Car in the 19th century \citep{Davidson1989}. Consequently, most LBV stars are surrounded by expanding structures of dust and gas. These circumstellar structures typically have very different shapes, from simple spherical shells to more complex, axisymmetric arrangements, including bipolar nebulae and rings \citep{Marston2006}. With characteristic times of $10^4$ years \citep{Humphreys1994}, the galactic population of LBV stars is extremely scarce, with less than twenty confirmed members \citep{Clark2005} and a few candidates which still require confirmation \citep{Stringfellow2012}. Circumstellar envelopes are also indirect probes of the existence of these sources, because they may contain information about past mass-loss events. \citep{Mizuno2010, Gvaramadze2012}.

LBV stars have been widely studied at optical and infrared wavelengths, with most of the research focusing on the determination of stellar parameters (temperatures, luminosities, metallicities and variabilities) and, to a lesser extent, the properties of the associated nebulae (neutral and ionized gas, and warm dust, \citealt{Umana2011, Ingallinera2014}). However, most of the underlying physical processes during the LBV phase are far from being totally understood. Little is known about the mechanisms that drive its variability and violent outbursts, and its actual role  in the late evolution of high mass stars is unclear, with recent observations suggesting that LBV stars may be (under certain conditions) direct progenitors of type II supernovae \citep{Groh2013}. For these reasons, the current picture of the LBV phase is rather incomplete. As many LBVs are surrounded by complex and multi-phase envelopes, targeting other components associated with these stars may provide valuable information to tackle some of these questions. In this regard, molecular gas is specially worth of attention. This component has been historically overlooked, as molecules were thought to be rapidly destroyed by the strong UV radiation in the outskirts of LBV stars. Despite of this, a series of pioneering studies have resulted in the detection of CO and NH$_3$ around a few LBVs. These detections prove that significant amounts of molecular gas can form and survive for long time-scales around sources of this type.

The best example is G79.29+0.46, a LBV nebula where \cite{Rizzo2008} reported nested shells of warm CO, strikingly attached to the infrared nebula and tracing successive mass-loss events. In a further study, \cite{Rizzo2014} found several NH$_3$ warm clumps clearly associated with the star, proving that the chemistry in the environs of LBV stars is more complex than initially assumed. These results underlined the potential of dynamical studies of molecular gas associated with LBV stars to achieve a complete vision of this elusive phase.

A particularly interesting but still unexplored case is that of MGE 042.0787+00.5084, a recently identified candidate LBV star. This source exhibits a nearly spherical circumstellar envelope with intense infrared emission, that was initially discovered in the \textit{Spitzer} MIPSGAL 24 $\mu$m survey \citep{Mizuno2010}. The central star was later catalogued as a candidate LBV by means of near-infrared spectroscopy, which revealed strong hydrogen emission lines (including Pfund and Brackett series) and several metal lines (\ion{Mg}{ii}, \ion{Na}{ii}, \ion{Fe}{ii}) usually found in other LBV spectra \citep{Flagey2014}. On the other hand, recent 6-cm VLA observations \citep{Ingallinera2016} unveiled a clumpy radio nebula strongly coupled to the infrared shell, with hints of a differentiated spectral index, as in other well known LBV sources such as HR Car \citep{Buemi2017}. 

In this paper we report the first detection of molecular material around MGE 042.0787+00.5084, using IRAM 30-m observations of CO and \co13 $J=1\rightarrow0$ and $J=2\rightarrow1$ lines. In Sect. \ref{sec:observations} the observing setup and procedures are detailed. In Sect. \ref{sec:results} we present the results of the observations. In Sect. \ref{sec:discussion} we estimate the physical parameters of the gas and provide an interpretation for the observed structure based on the derived mass-loss rate, which is compatible with the LBV nature of the object. We summarize our findings in Sect. \ref{sec:conclusions}.

\section{Observations} \label{sec:observations}

We observed MGE 042.0787+00.5084 with the 30-m radio telescope of the Institut de RadioAstronomie Millim\'etrique in Granada (Spain). Observations took place the night of 2017 July 24, under good summer conditions (opacity $\sim 0.25$, corresponding to 4.2 mm of precipitable water vapour at 225 GHz), with a total observing time of 3 hours.

Telescope was configured to operate in dual-receiver mode to map the distribution of CO and \co13 at 1.3 and 3 mm simultaneously, while covering other molecules of interest such as C$^{18}$O and C$^{17}$O. EMIR receivers E090 and E230 were used with two spectral setups: a first setup addressing CO $J=1\rightarrow0$, CO $J=2\rightarrow1$ and \co13 $J=1\rightarrow0$, and a complementary setup specifically targeting \co13 $J=2\rightarrow1$. The selected backend was the FTS spectrometer, providing an instantaneous bandwidth of 4 GHz per polarization and a velocity resolution of 0.26 and 0.52 \kms\ at 1 and 3 mm respectively.

On-the-fly (OTF) mapping in zig-zag mode was used to make $1\farcm5\times1\farcm5$ maps around the source, using a position-switching strategy with a common reference for all subscans. Two orthogonal mapping directions were used for each setup, with 21 parallel subscans per direction. Subscan length was set to 90\arcsec{} with a mapping speed of 3\arcsec{} s$^{-1}$, using a Nyquist-compliant spacing of 4\farcs3 between subscans.  The reference was located 10$\arcmin$ away of the source to minimize background/foreground contamination from the galactic plane. Map resolution corresponded to the half-power beam width of the telescope, which is $\sim11\arcsec$ and $\sim23\arcsec$ at 1 and 3 mm respectively. In addition, a deep integration towards the star's position was done to search for other, less abundant molecules.

We used Saturn and the quasar 1749+096 for pointing correction, and G34.3+0.2 for line calibration. Pointing was regularly checked, while calibrations were run every fifteen minutes and between consecutive maps. The pointing accuracy was always better than 5\arcsec.

\begin{table}
\centering
\caption{Observational parameters}
\label{tab:obs-parameters}
\begin{tabular}{lcccc}
\hline
Line & Freq & $\Delta$V  & HPBW & rms\\
& [GHz] & [\kms] & [$\arcsec$] & [mK] \\
\hline
CO $J=1\rightarrow0$ & 115.27 & 0.51 & 21.3 & 25 \\
\co13 $J=1\rightarrow0$ & 110.20 & 0.53 & 22.3 & 10 \\
CO $J=2\rightarrow1$ & 230.54 & 0.26 & 10.7 & 30 \\
\co13 $J=2\rightarrow1$ & 220.40 & 0.27 & 11.2 & 20 \\
\hline
\end{tabular}
\end{table}

A summary of the observational parameters for the target lines is provided in Table \ref{tab:obs-parameters}. Individual spectra were baseline-subtracted and combined to produce data cubes for each rotational transition. The data reduction and analysis was carried out with \textsc{gildas}, a software package especially tailored to deal with native 30-m data \citep{Guilloteau2000}. Hereafter in this work we adopt the following conventions: 
(1) data is presented in main beam temperature ($T_\mathrm{MB})$ scale unless specified otherwise. Conversion from antenna temperature ($T_\mathrm{A}^*$) is trivial via the relation $T_\mathrm{MB} = T_\mathrm{A}^*/\eta_\mathrm{eff}$, with 

\begin{equation}
\eta_\mathrm{eff} = \frac{\eta_\mathrm{MB}}{\eta_\mathrm{l}} \approx 0.9 \exp-\left( \frac{\nu\mathrm{[GHz]}}{399.7} \right) ^2
\end{equation}

where $\eta_\mathrm{MB}$ and $\eta_\mathrm{l}$ denote the antenna main beam and forward efficiencies respectively, expressed as a function of the frequency for the IRAM 30-m telescope \citep{SanchezContreras2015}; (2) velocities are expressed with respect to the local standard of rest  ($V_\mathrm{LSR}$); (3) positions are given as offsets relative to the J2000 equatorial coordinates of the source, which are (RA, Dec.) = (19\degr06\arcmin24\farcs57, +08\degr22\arcmin01\farcs9); and (4) position angles are measured from north to east.

\section{Results} \label{sec:results}

\begin{figure}
\includegraphics[width=\columnwidth]{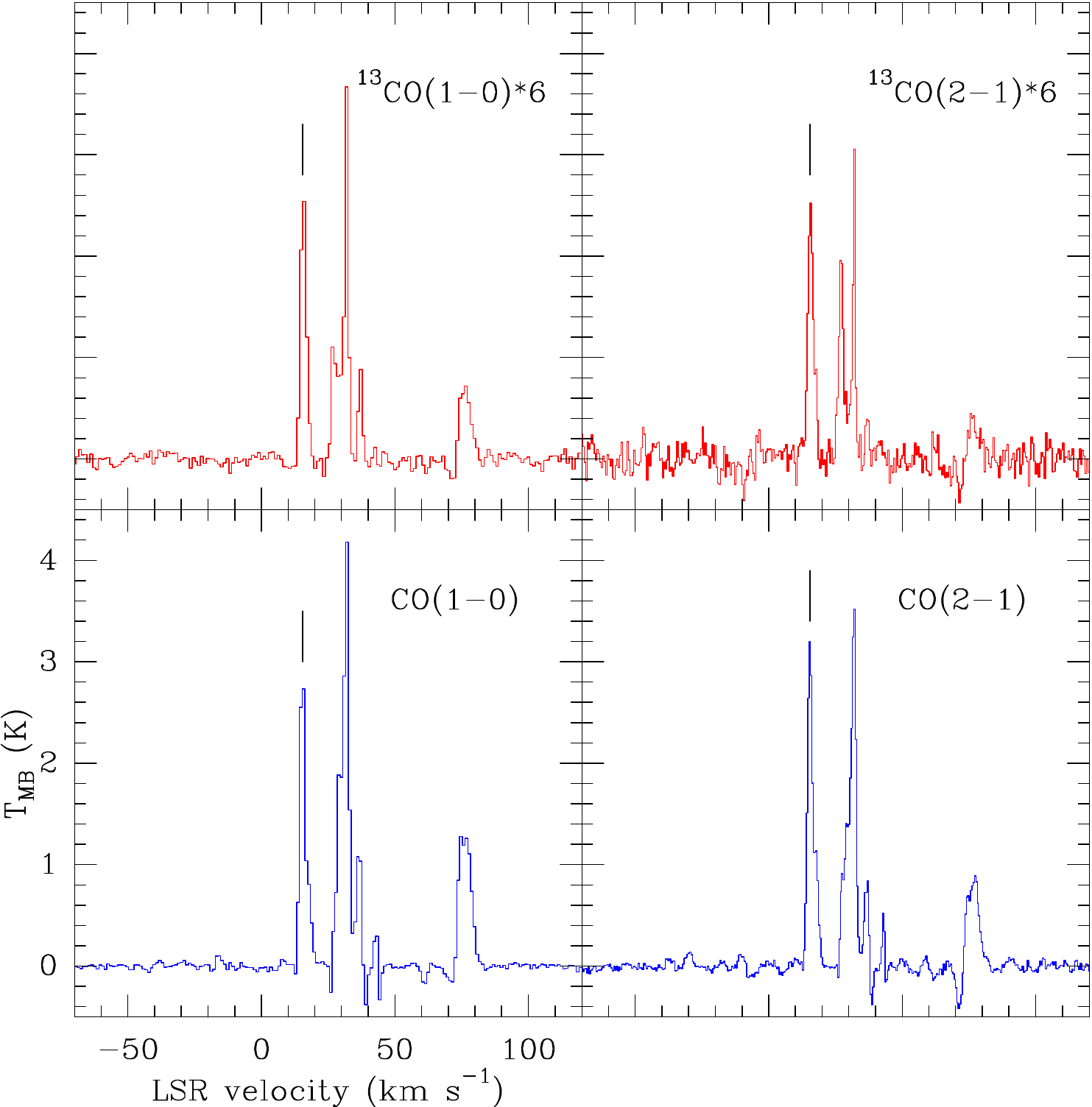}
	\caption{CO and \co13 spatially averaged spectra towards MGE042.0787+00.5084. \co13 spectra is scaled by a factor of 6 for easier comparison. Hanning smoothing has been applied to reduce the noise. The short vertical line on top of each spectra marks the component in the range (+13,+19)\kms .}
	\label{fig:rawspectra}
\end{figure}

After a thorough search in the whole data cubes we found significant emission of CO and \co13 at different velocities in the range from 10 to 80 \kms. Figure \ref{fig:rawspectra} displays the spectra of the four transitions spatially averaged over the whole field. C$^{18}$O was detected only towards the southeast of the source, and C$^{17}$O was not detected at any position.

\begin{figure*}
	\includegraphics[width=\textwidth]{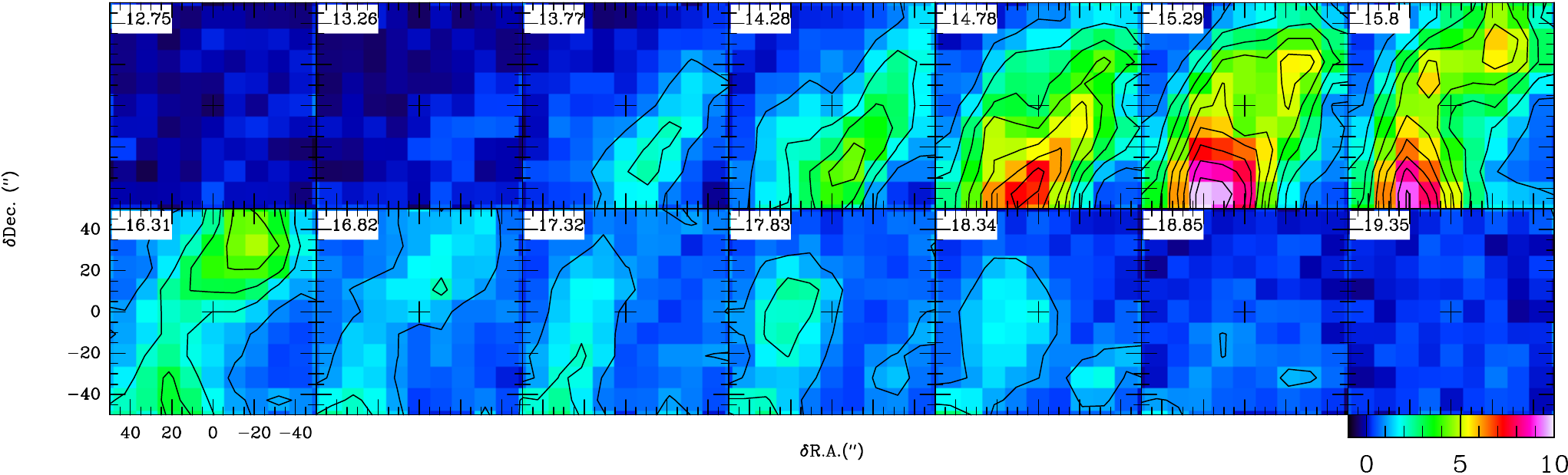}
	\caption{CO $J=1\rightarrow0$ emission towards MGE 042.0787+00.5084 in the velocity range 12.75--19.35 \kms. Absolute $V_\mathrm{LSR}$ is shown in the top left corner of each panel. Colour scale represents $T_\mathrm{MB}$ in K. Contours start at 1 K with a spacing of 1 K. Crosses at (0,0) indicate the position of the star.}
	\label{fig:chan_co10}
\end{figure*}

\begin{figure*}
	\includegraphics[width=\textwidth]{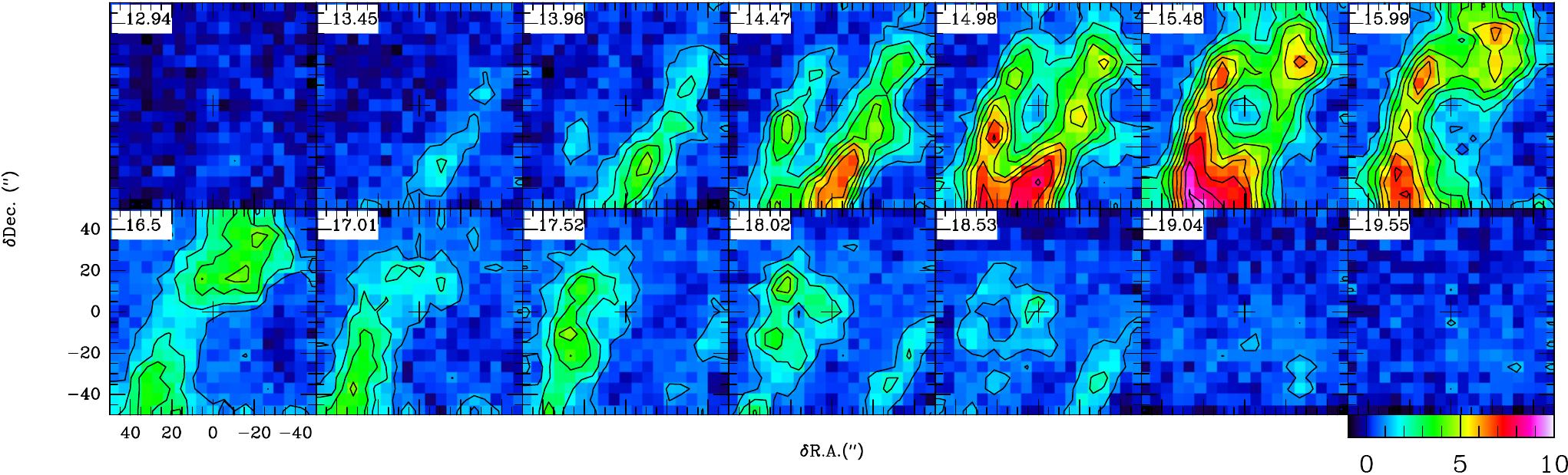}
	\caption{Same as Fig. \ref{fig:chan_co10} for CO $J=2\rightarrow1$, with a channel binning of 2 (resolution $\sim$ 0.5 \kms) for easier comparison. Contours start at 1 K with a spacing of 1 K. }
	\label{fig:chan_co21}
\end{figure*}

\begin{figure*}
	\includegraphics[width=\textwidth]{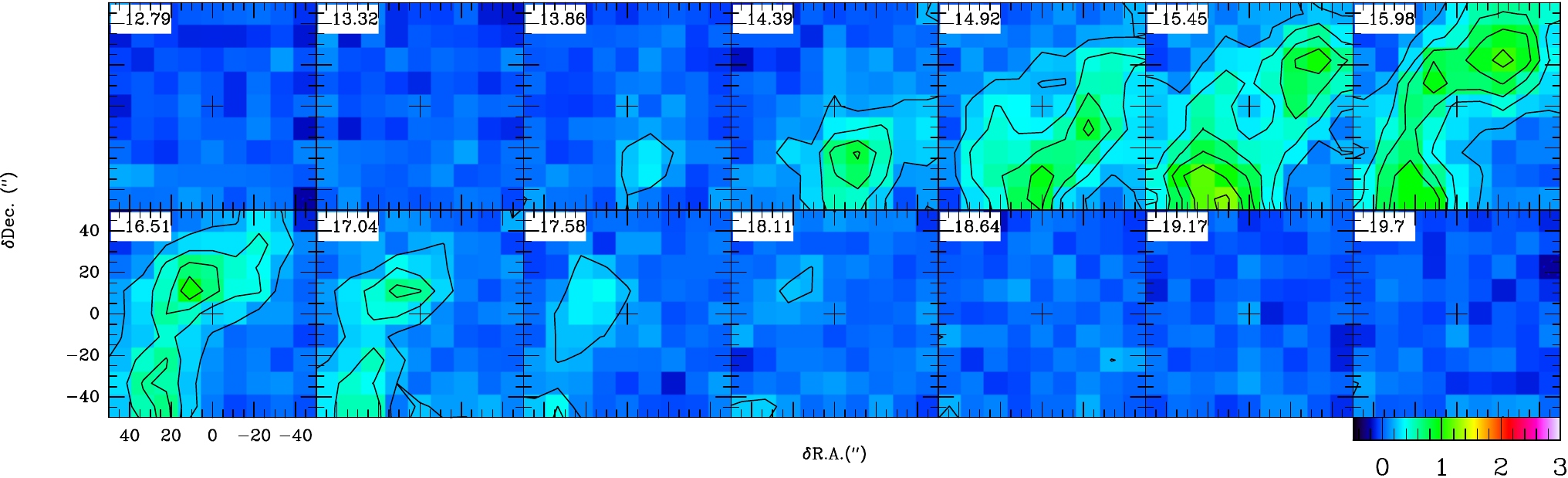}
	\caption{Same as Fig. \ref{fig:chan_co10} but for \co13 $J=1\rightarrow0$. Contours start at 100 mK with a spacing of 100 mK. }
	\label{fig:chan_13co10}
\end{figure*}

\begin{figure*}
	\includegraphics[width=\textwidth]{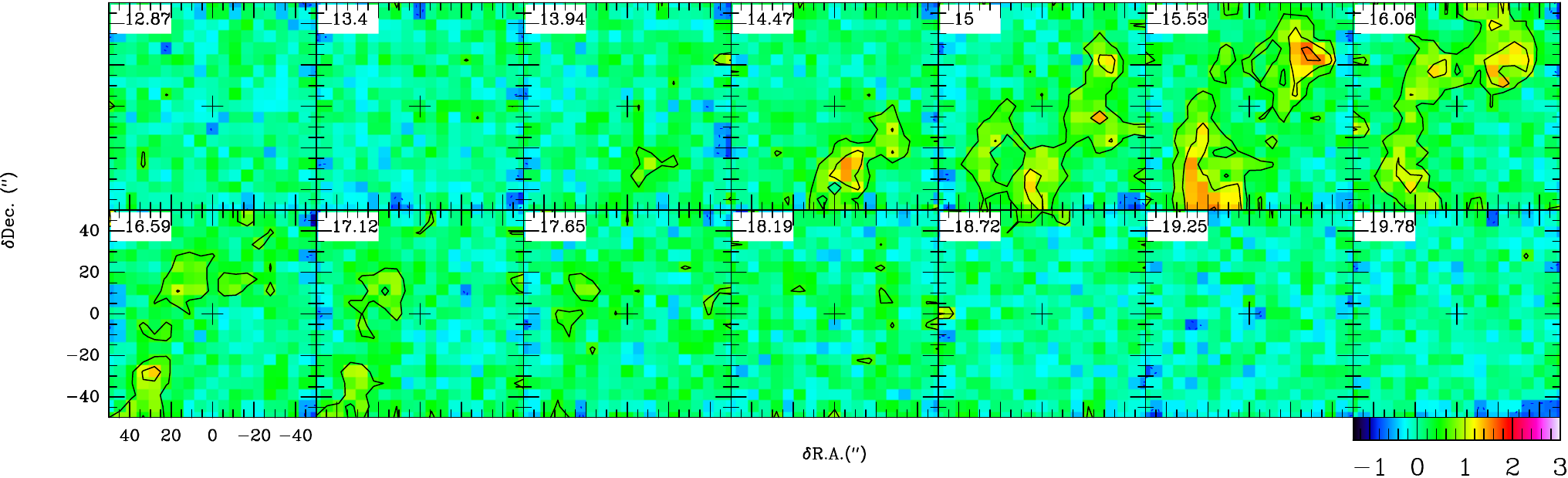}
	\caption{Same as Fig. \ref{fig:chan_co10} but for \co13 $J=2\rightarrow1$, with a channel binning of 2 (resolution $\sim$ 0.5 \kms) for easier comparison. Contours start at 200 mK with a spacing of 200 mK. }
	\label{fig:chan_13co21}
\end{figure*}

A detailed analysis of the spatial distribution of the emission reveals that, in the four CO and \co13 lines, most of the velocity components are present over a significant part of the observed field, without a clear pattern with respect to the star; these components are presumably not related to the source, being most likely foreground/background contamination arising from molecular clouds in the line of sight. In contrast, the velocity component in the range (+13,+19) \kms, marked with vertical lines in the spectra, depicts certain symmetry around the central star, in the form of a rather elongated ring-like structure. This component is represented as channel maps in Figs. \ref{fig:chan_co10} to \ref{fig:chan_13co21}, in steps of $\sim$ 0.5 \kms. Several arcs more or less centred at the star position are noted; these arcs present different sizes at different velocities, moving from southwest to northeast as velocity increases. At $\sim$15--15.5 \kms\, the emission is closed in elliptical features, especially noted in the CO $J=2\rightarrow1$ line.

\begin{figure}
	\includegraphics[width=\columnwidth]{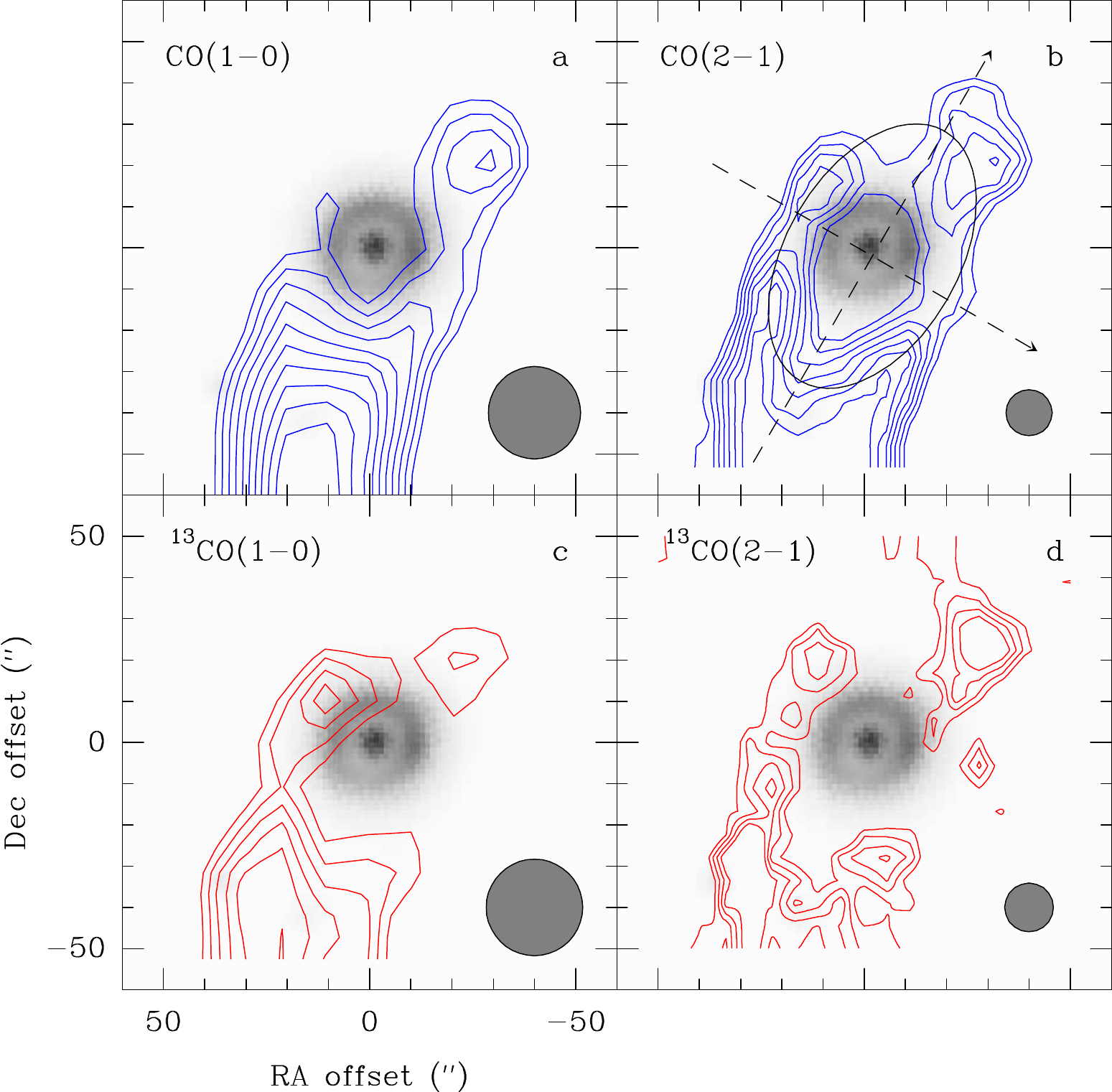}
	\caption{Velocity-integrated maps of CO and \co13 towards MGE042.0787+00.5084 in the range (+13.4,+17.4) \kms, shown as contours, superimposed on the \textit{Spitzer} MIPSGAL 24 $\mu$m image of the infrared nebula \citep{Ingallinera2016}. Contour starting values and steps are 6.5 and 0.6; 6.5 and 1.2; 1.35 and 0.15; and 1.5 and 0.25 K \kms\, for panels a, b, c and d respectively. The grey circle in the bottom right corner of each panel represents the beam size. Dashed lines in panel b indicate the cuts for the position-velocity diagrams in Fig. \ref{fig:pv-cuts}, and the ellipse used for the fitting is shown for reference.}
	\label{fig:intensity}
\end{figure}

Figure \ref{fig:intensity} shows the corresponding velocity-integrated intensity maps, as contours superimposed on the 24 $\mu$m \textit{Spitzer} MIPSGAL image of the infrared nebula (taken from \citealt{Mizuno2010}). These maps reveal a rather clumpy ring-like structure centred at the star position, displaying a remarkable elongation in the approximate southeast-northwest direction. By fitting an ellipse to the emission peaks as shown in the figure, we derive an angular size of $35\arcsec\times21\arcsec$ with a position angle of 150$\degr$.

The structure is mostly unresolved so we are not able to determine its actual thickness. Never the less, the 30-m angular resolution is sufficient to observe a certain degree of morphological differentiation, allowing us to identify a number of regions that might deserve further attention:

\begin{enumerate}
\item The main ring-like feature, which completely surrounds the infrared nebula and accounts for most of the emission.
\item The inner cavity, a region in which molecular material is likely depleted. The extent of this cavity strikingly matches the infared nebula.
\item The partially isolated emission blob towards the northwest, hereafter the northwestern clump, projected $\sim25''$ from the star. This region is particularly prominent in the $J=2\rightarrow1$ images.
\item The southeast region, from which arises the most intense and widespread emission.
\end{enumerate}

\begin{figure}
\includegraphics[width=\columnwidth]{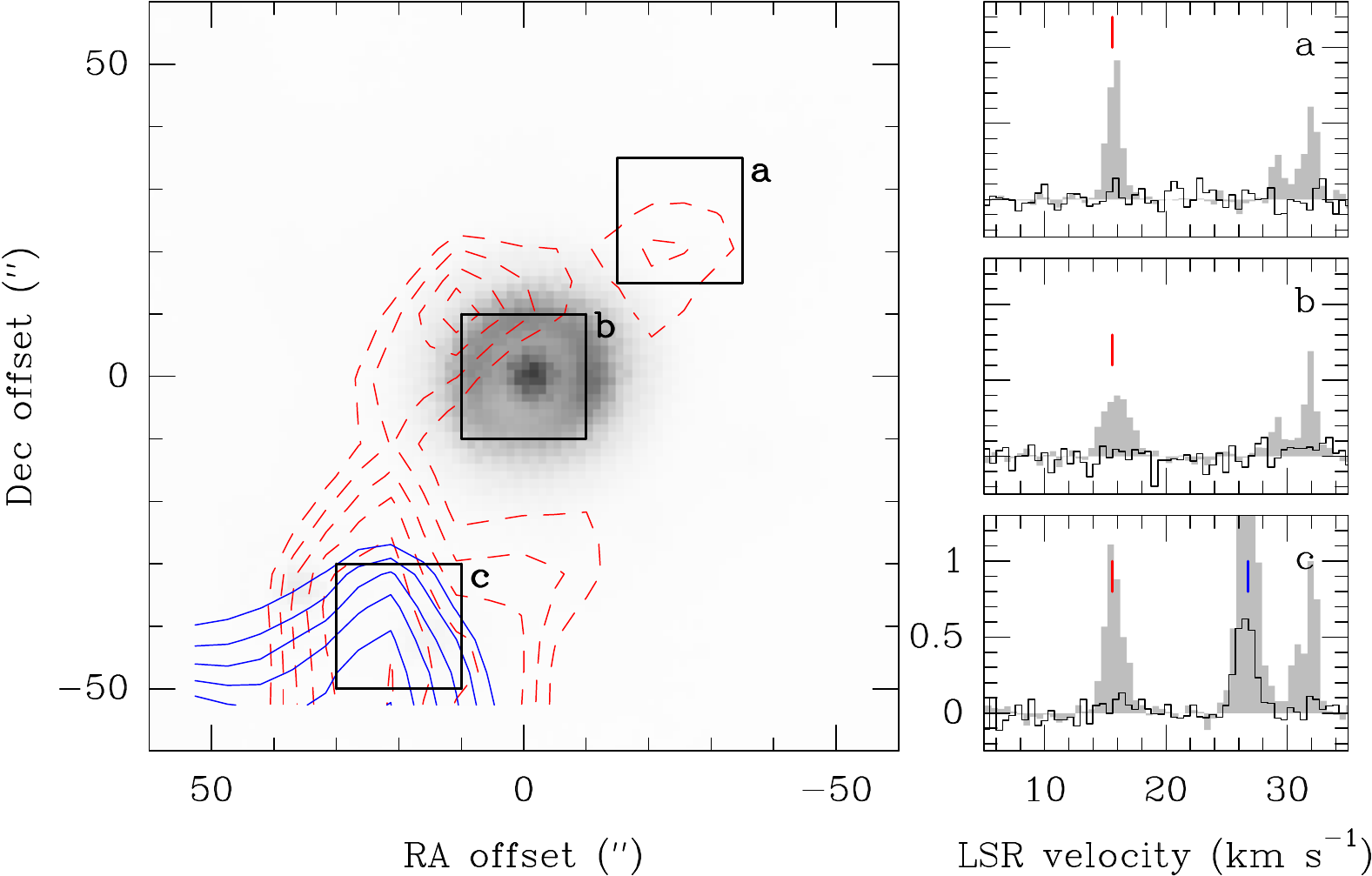}
	\caption{Overabundance of \co13. Left panel: Contour maps representing the emission of C$^{18}$O $J=1\rightarrow0$ integrated in the velocity range (+25.6,+27.8)\kms\ (solid) and \co13 $J=1\rightarrow0$ in the velocity range (+13.4,+17.4)\kms\ (dashed), superimposed on the \textit{Spitzer} MIPSGAL 24 $\mu$m image of the infrared nebula in gray scale. Contours starting values and steps are 0.4 and 0.1 and 1.35 and 0.15 respectively, in units of K\kms. The labelled squares represent regions of $20\arcsec\times20\arcsec$ where the spectra in the corresponding right panels have been spatially averaged. Right panels: averaged spectra of C$^{18}$O $J=1\rightarrow0$ (black) and \co13 $J=1\rightarrow0$ (filled) over the region with the corresponding label. The former has been scaled by 2 for easier comparison. The velocity components represented by the contour maps are marked with vertical lines. Intensity scale in units of K.}
	\label{fig:c18o}
\end{figure}

This southeast region, where the overall symmetry of the structure is lost, may be seriously contaminated by emission from a molecular cloud at a close $V_\mathrm{LSR}$. Figure \ref{fig:c18o} compares the spectra of \co13 and C$^{18}$O $J=1\rightarrow0$ averaged over three different parts of the map: the clump, the stellar position and the southeast region. C$^{18}$O is only detected in the latter, as a weak and narrow component in the velocity range (+25.6,+27.8)\kms. The non-detection of C$^{18}$O in other positions suggests the existence of a chemical differentiation, indicating that this widespread emission may be --at least partially-- disaggregated from the main ring-like structure.

\begin{figure}
\includegraphics[width=\columnwidth]{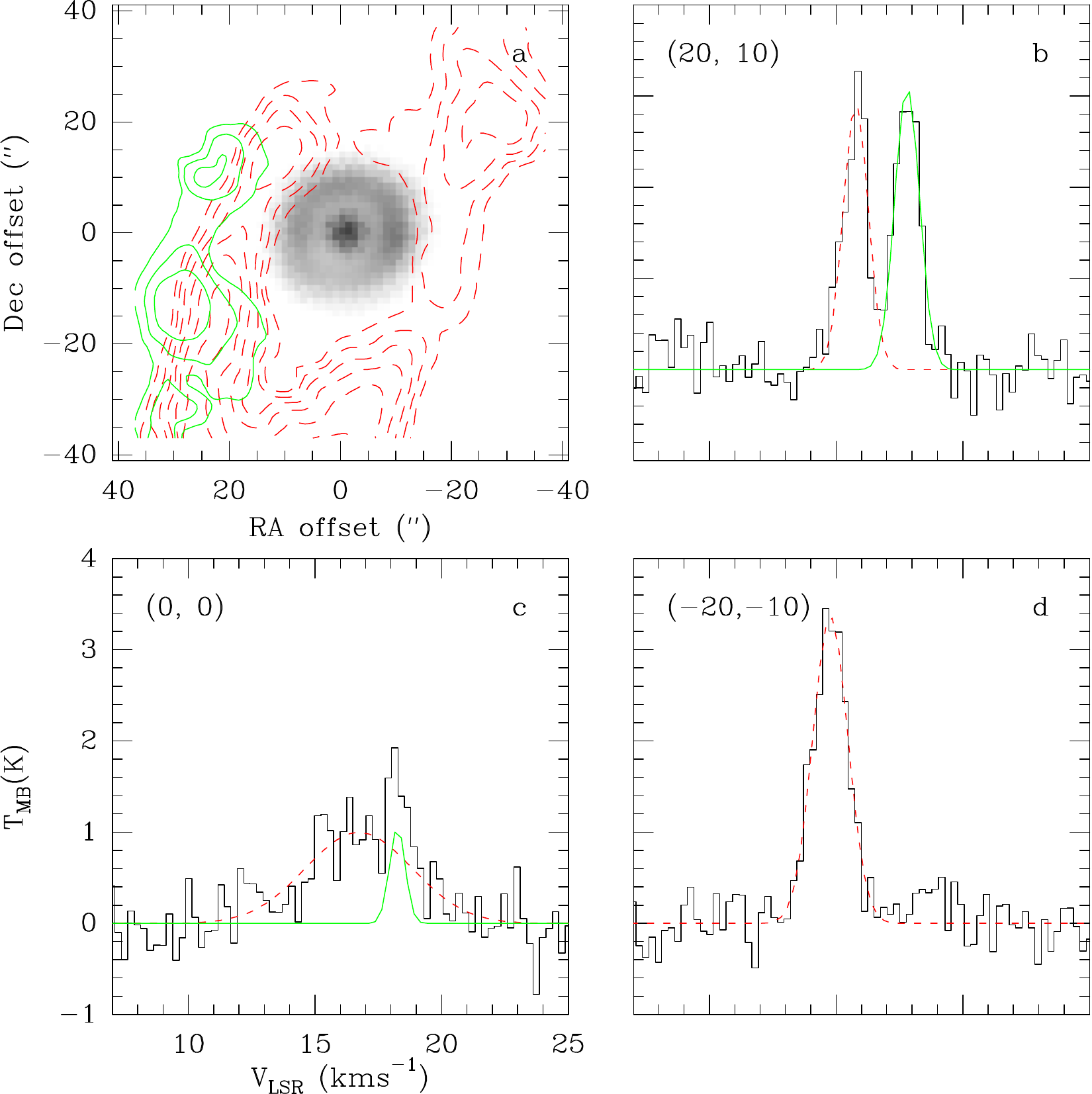}
	\caption{Hints of substructure in the CO gas: (a) the two velocity components as contours superimposed to the 24 $\mu$m image of the infrared nebula in gray scale. In dashed contours, the component at $V_\mathrm{LSR}$  = 15.4 \kms, with levels starting at 6.5 in steps of 1.2 K\kms; in solid contours, the component at $V_\mathrm{LSR}$ = 17.8 \kms, with levels starting at 2.5 with steps of 0.6 K\kms. (b) spectra at position (20\arcsec, 10\arcsec). (c) spectra at star's position (0\arcsec, 0\arcsec). (d) spectra at position (-20\arcsec, -10\arcsec). A gaussian fitting for each line is plotted. }
	\label{fig:components}
\end{figure}

As shown in Fig. \ref{fig:components}, a closer inspection of spectra in individual positions reveals that this structure is the blending of two different components at $V_\mathrm{LSR}$ = 15.4 and 17.8 \kms, depicted in dashed and solid contours respectively. The component at 15.4 \kms\ is relatively ubiquitous and completely surrounds the source in all directions, but the component at 17.8 \kms\ is highly localized and only visible in the CO lines. It is prominent towards the East, with a peak temperature comparable to the other component, but almost non-existent in other positions (e.g. southwest). Similarly, its motion with respect to the star is unclear. This makes extremely difficult to confirm whether this component is related to the source in any way: it could be part of an interfering cloud or a mere stratification of the main component. Regardless of its nature, the contribution of this component is negligible, as the overall emission is dominated by the component at 15.4 \kms. For this reason, we limit our analysis to the latter, adopting its central velocity as the systemic velocity of the structure.

\begin{figure*}
	\includegraphics[width=\textwidth]{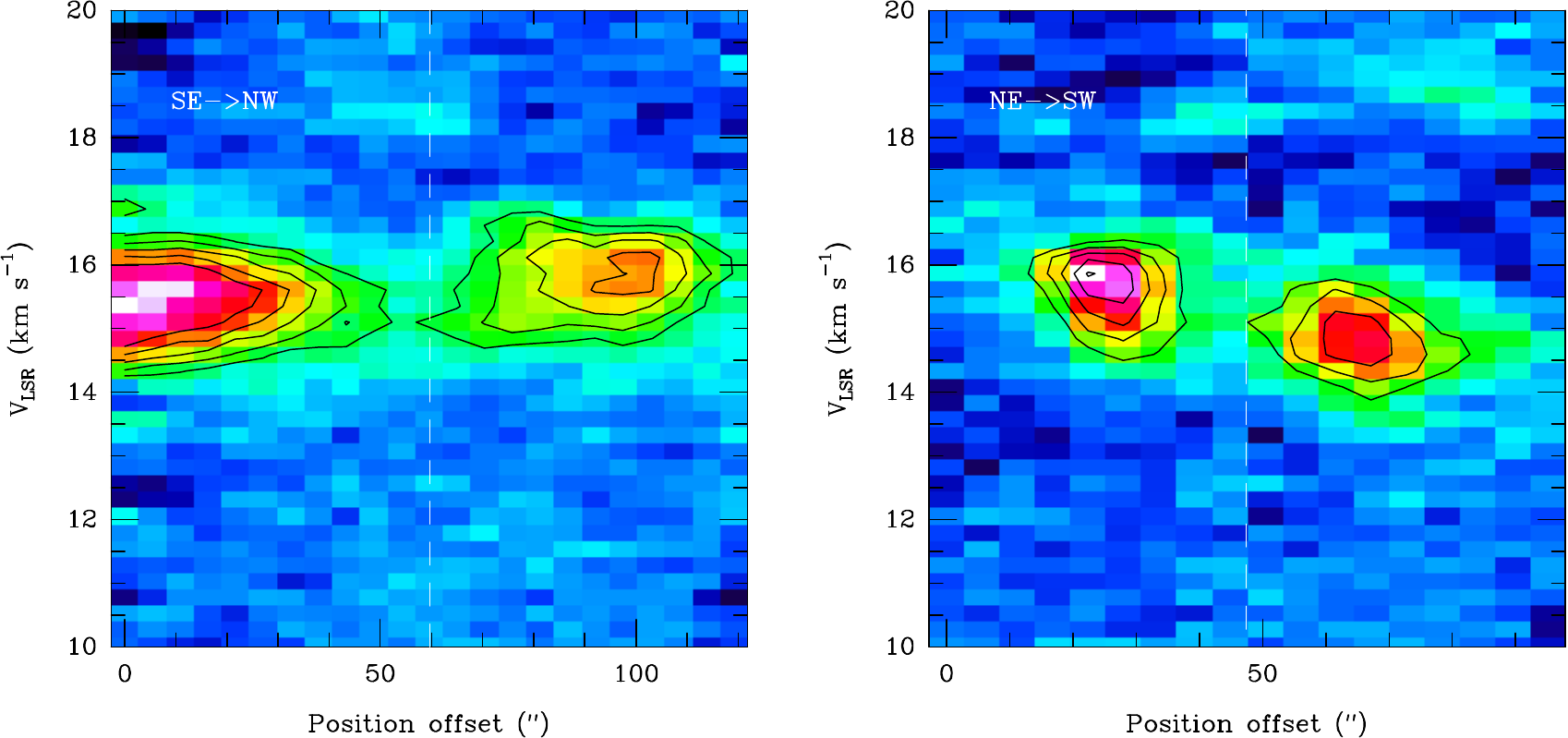}
	\caption{Position velocity diagrams of the CO $J = 2 \rightarrow1$ line in the slices defined in Fig. \ref{fig:intensity}. The direction of the cut is indicated in the top left corner of each panel, and the vertical dashed line represents the position of the central star.}
	\label{fig:pv-cuts}
\end{figure*}

For a better understanding of the velocity structure of the CO and \co13 emission, we studied the kinematics of the gas by means of position-velocity (PV) diagrams along the strips indicated with dashed lines in Fig. \ref{fig:intensity}, which were selected according to the axes of the ellipse previously used for the fitting. The first slice goes from SE to NW (P.A. 150\degr), following the direction of maximum elongation, and the second is orthogonal, from NE to SW (P.A. 60\degr). As seen in Fig. \ref{fig:pv-cuts}, we find emission highly concentrated in two isolated blobs either side of the star, each of them spanning for $\sim$ 2 \kms. The velocity difference between the blobs is small regardless of the direction of the cut, in the range 0.5--1.5 \kms\ from peak to peak.

\begin{figure}
	\includegraphics[width=\columnwidth]{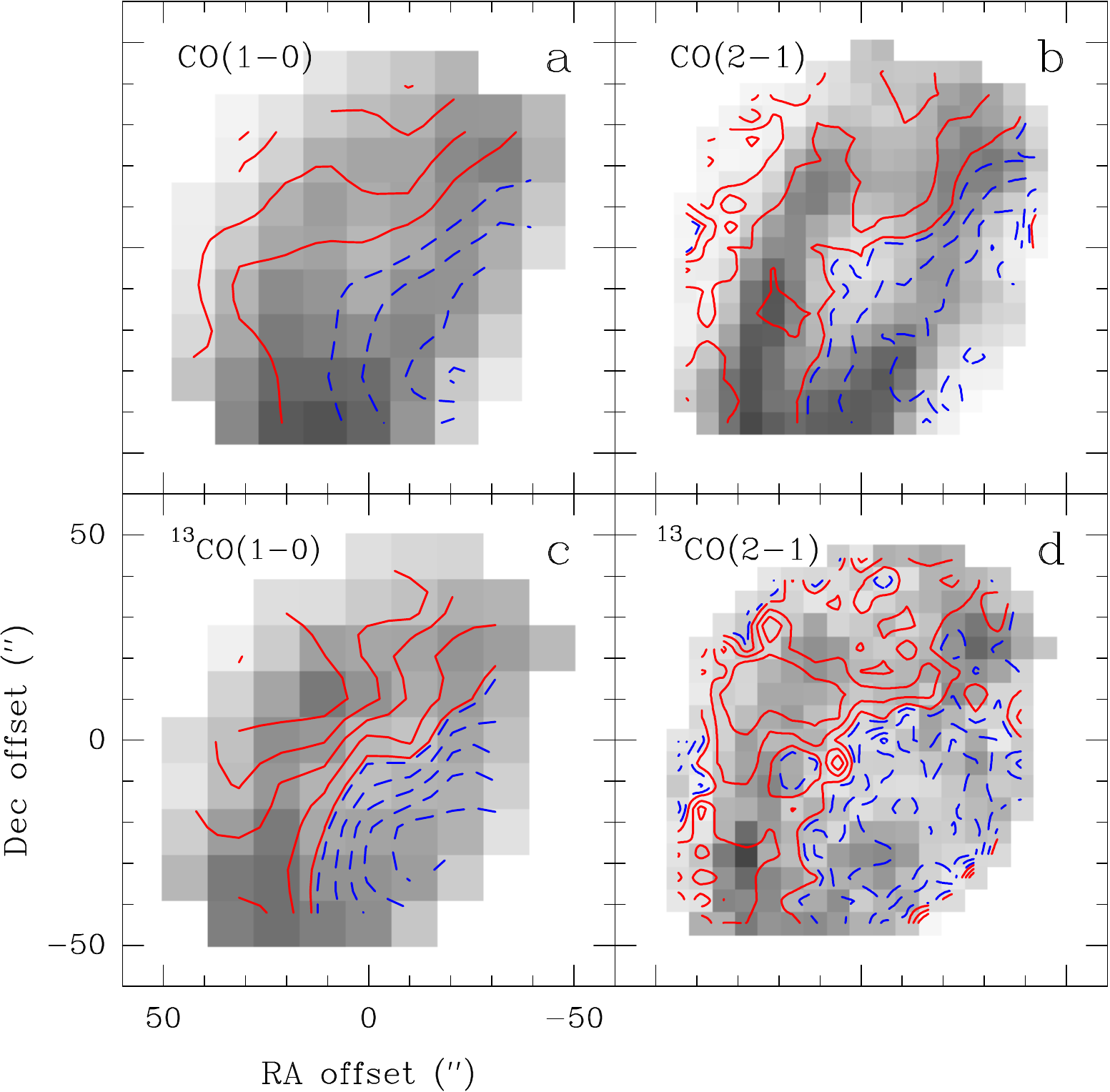}
	\caption{Intensity-weighted velocity maps of CO and \co13 towards MGE042.0787+00.5084 shown as contours. Gray scale represents the intensity of each transition at the original resolution, in units of K \kms. Dashed blue contours indicate $V_\mathrm{LSR} < V_0$ and solid red contours $V_\mathrm{LSR} > V_0$. Contour spacing is 0.15 and 0.25 K \kms for $J=1\rightarrow0$ and $J=2\rightarrow1$ lines respectively. Border of the maps have been masked out to minimise noise.}
	\label{fig:velocity}
\end{figure}

\section {Discussion} \label{sec:discussion}

\subsection{Morphology and kinematics} \label{sec:discussion:morfkin}

Despite the contribution from the southern cloud, most of the emission arises from a circumstellar ring-like feature that encloses the infrared shell. The PV diagrams in Fig. \ref{fig:pv-cuts} showed that material in the ring moves at different velocities either side of the exciting star, but a deeper analysis is required to disentangle the motion of the structure. We resorted to study the first-order moment of the emission, i.e. the intensity-weighted velocity distribution, presented in Fig. \ref{fig:velocity}, which approximates the velocity field of the gas. The four CO and \co13 lines share a common pattern, a smooth velocity gradient in the NE-SW direction that roughly splits the field across the semimajor axis of the structure. Therefore, the northeast and southwest regions show opposite motions, represented by two sets of contours in the maps: gas in the northeast moves away (redshifted emission) while gas in the southwest moves towards the observer (blueshifted emission).

The spatial distribution of the emission can be explained by a torus seen at a certain angle, but the velocity gradient observed is compatible with a structure undergoing expansion or contraction. However, near-infrared spectroscopic observations by \cite{Flagey2014} proved that MGE 042.0787+00.5084 is a massive evolved star, allowing us to confidently rule out the contraction scenario --i.e. material collapsing/accreting onto the star--, and therefore conclude that the gas is distributed in an expanding toroidal structure. 

If the ring is expanding isotropically in a uniform medium, the observed elongation may be interpreted as a mere projection effect, which allows us to infer the inclination angle by fitting ellipses to the emission maxima as described in Sect. {\ref{sec:results}},  and calculating the minor to major axis ratios. With this method we derive an inclination of $53\pm6\degr$. However, the overall morphology of the emission is heavily altered by the southern widespread component, and a local density gradient in the direction of the major axis may also contribute to the elongation, so we could be overestimating the inclination of the structure.

In this regard, we note similar velocity shifts in the two PV diagrams from Fig. \ref{fig:pv-cuts}, showing just a slightly larger difference in the NE-SW cut, which is likely more affected by projection according to the viewing geometry. This could be taken as an indicator of a low inclination. Therefore we cannot merely rely on the ellipse fitting to constrain the inclination angle of the structure, and in Sect. \ref{sec:modelresults} we considered a larger range of possible inclinations.  

We can use the NE-SW cut to measure the projected expansion velocity as half the velocity difference between the emission blobs. We estimate a projected velocity of $\sim1$\kms, which relates to the expansion velocity as

\begin{equation}
V_\mathrm{exp} = \frac{\Delta V}{2 \sin i},
\end{equation}

\noindent where  $i$ is the inclination angle and $\Delta V$ the increment in velocity. Using the previously estimated inclination angle, we obtain $V_\mathrm{exp} =$ 1.2--1.4 \kms. This is an extremely low expansion velocity, an order of magnitude lower than those measured in other LBV sources (e.g. the main CO shell in G79.29+0.46 expands at 14 \kms, \citealt{Rizzo2008}), but it is important to bear in mind that this result is just a coarse lower limit, strongly affected by the inclination angle. Lower inclinations would be compatible with higher velocities. 

\subsection{Physical parameters of the gas}

Being MGE 042.0787+00.5084 a high-mass evolved star, mass-loss processes --both heavy steady winds and/or eruptive mass ejections-- could have originated the structure seen in CO and \co13. The simultaneous observation of two rotational lines from two isotopologues allows us to derive some physical parameters of the gas. 

\begin{figure}
\includegraphics[width=\columnwidth]{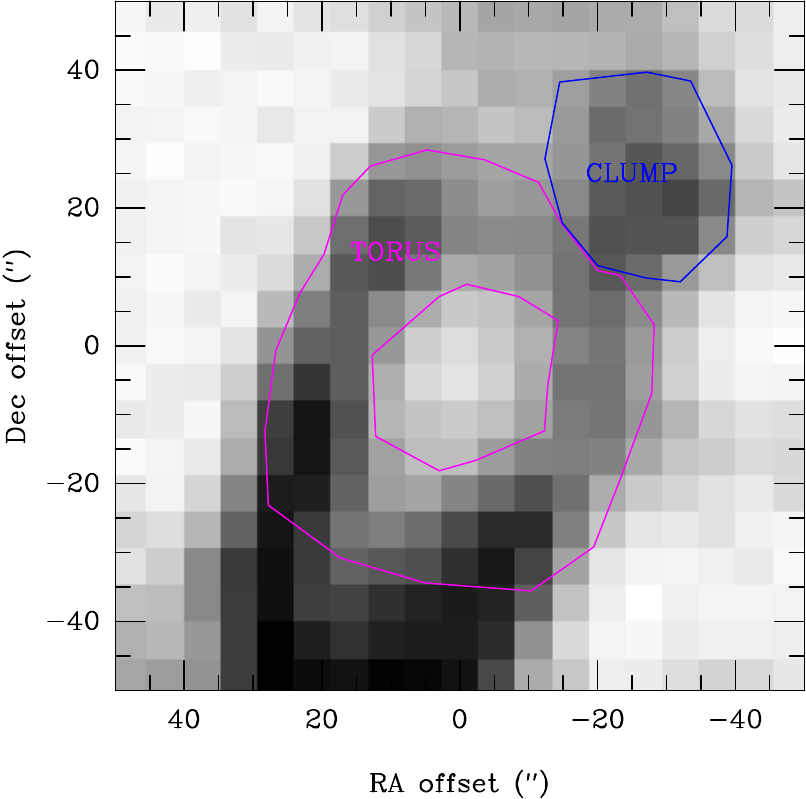}
	\caption{Regions in which physical parameters were derived independently: the northwestern clump (blue) and the main torus (magenta). For reference, CO $J=2\rightarrow1$ emission in the velocity range (+13.4, +17.4) \kms\ is shown in greyscale.}
	\label{fig:regions}
\end{figure}

To do so, we performed LTE (local thermodynamic equilibrium) and non-LTE analyses in two separate regions, to account for the spatial differentiation observed in the structure: the torus as a whole, and the northwestern clump. The two regions are depicted in Fig. \ref{fig:regions}. We excluded most of the the southeastern region to minimize the contribution from the widespread emission.

\subsubsection{LTE analysis}
\label{sec:dis:lte}

As a first approach we determined the global physical parameters of the gas in the LTE scenario, using \co13, the least intense and presumably least abundant observed species. By assuming optically thin emission, the column density in the lower level is given by

\begin{equation}
N_\mathrm{l} = 1.94\times10^3 \nu^2 \frac{g_\mathrm{l}}{g_\mathrm{u}}\frac{1}{A_\mathrm{ul}} \int{T_\mathrm{MB} \mathrm{d}v},
\end{equation}

\noindent where $\nu$ is the frequency of the line, $A_\mathrm{ul}$ is the spontaneous decay rate, $g_\mathrm{J} = 2\mathrm{J}+1$ is the degeneracy of each rotational level and the integral term depends on the excitation temperature $T_\mathrm{ex}$ through the relation $\int{T_\mathrm{MB} \mathrm{d}v} = T_\mathrm{ex}\int{\tau \mathrm{d}v}$, which is true in the optically thin regime \citep{Wilson2009}. In LTE, levels are populated following a Boltzmann distribution described by a single $T_\mathrm{ex}$. In order to relate $N_\mathrm{l}$ to the total population of all energy levels in the molecule we use a rotational partition function $Q(T_\mathrm{ex})$ such that,

\begin{equation}
N = 1.94\times10^3 \nu^2 \frac{g_\mathrm{l}}{g_\mathrm{u}}\frac{Q(T_\mathrm{ex})}{A_\mathrm{ul}}  \exp{\bigg(\frac{E_\mathrm{u}}{kT_\mathrm{ex}}\bigg)}  \int{T_\mathrm{MB} \mathrm{d}v},
\end{equation}

\noindent where $E_\mathrm{u}$ is the energy of the upper level and $T_\mathrm{ex}$ is the excitation temperature. The latter can be estimated from the $2\rightarrow1 / 1\rightarrow0$ line ratio by solving the relation

\begin{equation}
\frac{T_\mathrm{R}(\mathrm{21})}{T_\mathrm{R}(\mathrm{10})} = 4\exp{\bigg(-E_{21}/T_\mathrm{ex}\bigg)} ,
\label{eq:coldens-lte}
\end{equation}

\noindent with $T_\mathrm{R}(\mathrm{ul})$ the peak temperature of the $\mathrm{ul}$ transition. The derived excitation temperatures --between 9 and 12 K-- and column densities -- all in the range of $10^{15}$ cm$^{-2}$-- are summarized in Table \ref{tab:cd-lte}. The column densities of each molecule do not vary significantly between the torus and the clump.

\begin{table*}
\caption{LTE column density estimates.}
\label{tab:cd-lte}
\begin{tabular}{ccccc}
\hline
& $T_\mathrm{ex}$(\co13) & N(\co13) & $T_\mathrm{ex}$(CO) & N(CO) \\
& $[\mathrm{K}]$ &  [10$^{15}$ cm$^{-2}$] & $[\mathrm{K}]$ & [10$^{15}$ cm$^{-2}$] \\
\hline
\textbf{Clump} & 11.2 & 1.0 (0.2) & 11.6 & 6.1 (0.6) \\
\textbf{Torus} & 9.8 & 1.0 (0.2) & 12.1 & 6.1 (1.1) \\
\hline
\end{tabular}
\end{table*}

\subsubsection{Non-LTE analysis}\label{sec:discussion:radex}

To study a more realistic scenario we ran simulations with the non-LTE excitation and radiative transfer code RADEX \citep{Tak2007}. RADEX uses the escape probability formulation with the Large Velocity Gradient (LVG) approximation \citep{Sobolev1960} to constrain the excitation conditions of the gas according to the observed line intensities. We involved the four lines in the fitting, convolving the $2\rightarrow1$ maps to the $1\rightarrow0$ resolution to obtain true line ratios.

The calculations were made taking the CO and \co13 collisional coefficients from LAMDA database \citep{Schoier2005}, with a background temperature of 2.73 K. First, we ran RADEX for a range of kinetic temperatures ($T_\mathrm{k}$) from 10 to 250 K in steps of 10 K, solving the radiative transfer in a two-parameter space defined by $n(\mathrm{H}_2)$ --the H$_2$ volume density--  and $N(^{13}\mathrm{CO})$, with the aim of finding the best fit for the observed \co13 $J=1\rightarrow0$ intensities and $(2\rightarrow1)/(1\rightarrow0)$ ratios. Later we cross-checked the resulting $n_\mathrm{H_2}$ values with the CO line intensities and ratios in order to determine $N$(CO) and restrict the range of possible $T_\mathrm{k}$.

\begin{table*}
\caption{RADEX results.}
\label{tab:cd-radex}
\begin{tabular}{cccccccc}
\hline
\multicolumn{8}{c}{\textbf{Clump}} \\
T$_\mathrm{K}$ & $n(\mathrm{H}_2)$ &  N(\co13)  & $\tau_{10}$ & $\tau_{21}$ &N(CO)  & $\tau_{10}$ & $\tau_{21}$\\
$[\mathrm{K}]$& [$10^3$ cm$^{-3}$]  &  [10$^{15}$ cm$^{-2}$]& & & [10$^{15}$ cm$^{-2}$] & &\\
\hline
10 & 7.2 (1.4) & 1.2 (0.1) & 0.05 & 0.11 & 8.3 (1.2) & 0.35 & 0.7 \\
30 & 1.5 (0.2) & 1.0 (0.2) & 0.02 & 0.09 & 7.1 (0.9) & 0.12 & 0.51 \\
50 & 1.0 (0.1) & 1.0 (0.2) & 0.01 & 0.09 & 6.8 (1.0) & 0.09 & 0.49 \\
70 & 0.8 (0.1) & 1.0 (0.2) & 0.01 & 0.08 & 6.8 (1.0) & 0.08 & 0.49 \\
\hline
\multicolumn{8}{c}{\textbf{Torus}} \\
T$_\mathrm{K}$ & $n(\mathrm{H}_2)$ &  N(\co13)  & $\tau_{10}$ & $\tau_{21}$ &N(CO)  & $\tau_{10}$ & $\tau_{21}$\\
$[\mathrm{K}]$& [$10^3$ cm$^{-3}$]  &  [10$^{15}$ cm$^{-2}$]& & & [10$^{15}$ cm$^{-2}$] & &\\
\hline
10 & 4.1 (1.2) & 1.3 (0.2) & 0.06 & 0.14 & 7.6 (1.7) & 0.39 & 0.71 \\
30 & 1.0 (0.2) & 1.2 (0.2) & 0.04 & 0.12 & 6.2 (1.8) & 0.19 & 0.56 \\
50 & 0.7 (0.2) & 1.1 (0.2) & 0.03 & 0.12 & 6.2 (1.8) & 0.18 & 0.56 \\
70 & 0.5 (0.1) & 1.1 (0.2) & 0.03 & 0.12 & 6.2 (1.5) & 0.17 & 0.55 \\
\hline
\end{tabular}
\end{table*}

\begin{figure*}
	\includegraphics[width=\textwidth]{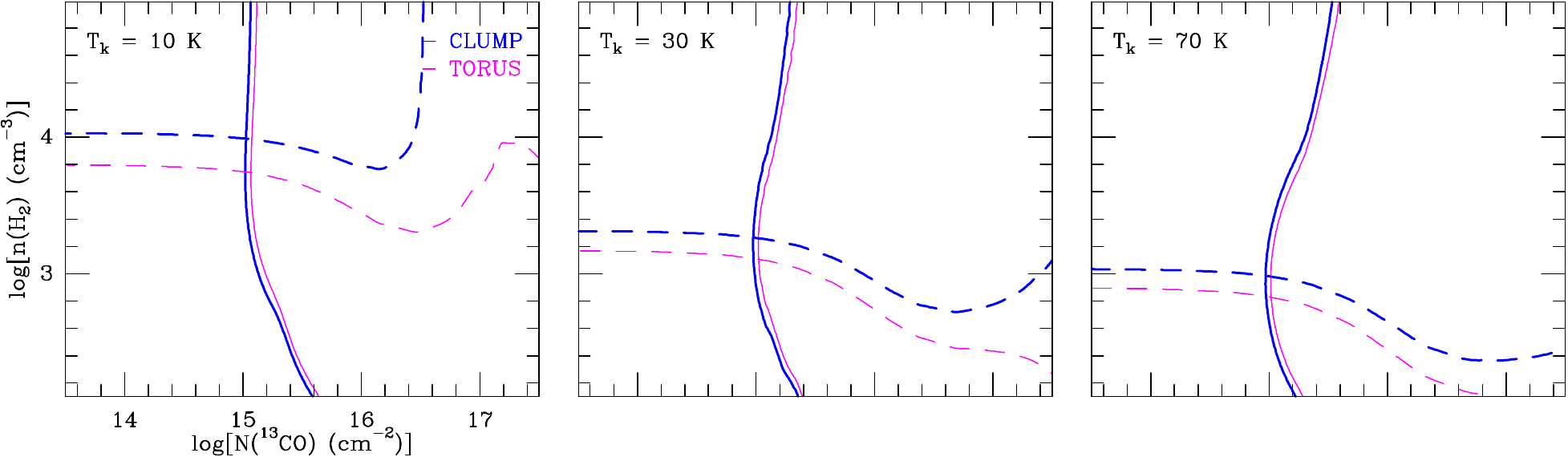}
	\caption{\co13 RADEX fittings computed for $T_\mathrm{k} = 10$ K (left), $T_\mathrm{k} = 30$ K (centre) and $T_\mathrm{k} = 70$ K (right), in the clump (thick blue line) and the torus (thin magenta line); see Fig. \ref{fig:regions}. Solid curves represent the observed \co13 $J=1\rightarrow0$ intensity, and dashed curves the \co13 ($2\rightarrow1$)/($1\rightarrow0$) ratio. Intersections between two sets of contours indicate the most likely values of $n(\mathrm{H_2})$ and  $N(^{13}\mathrm{CO})$.}
	\label{fig:lvg}
\end{figure*}

Simulation outcomes are compiled in Table \ref{tab:cd-radex} and some RADEX runs are presented in Fig. \ref{fig:lvg}. We find \co13 column densities in the range 1.0--1.3$\times10^{15}$ cm$^{-2}$, which only differ by 10--20\% from those of the LTE analysis. There is little variation between the column densities of the clump and the torus,  with \co13 opacities supporting the optically thin assumption. Similarly, the \co13 and CO column densities are relatively insensitive to temperature, which is consistent with optically thin emission as well. Considering the uncertainties involved, $N$(CO) is only 5--7 times higher than $N$(\co13), which indicates a surprisingly low isotopic ratio. The implications of this finding are discussed in Sect. \ref{sec:isotopic}.

The resulting gas densities are quite low, of a few 10$^3$ cm$^{-3}$, and we observe significant differences between the torus and the clump, with the latter being 1.5--2 times denser. These low densities are consistent with the non-detection of other high-density tracers, and somehow limit the range of possible kinetic temperatures, with solutions up to 70 K for the torus and 100 K for the clump. Above these values, $n_\mathrm{H_2}$ falls well below the critical density of the $1\rightarrow0$ transition. 

In the case of CO we find higher $\tau$ values, suggesting that the observed intensities may be slightly affected by opacity, specially at low temperatures. In addition, we observe a significant dilution in the clump after convolution, which contrasts with the prominence of this region in the original maps. This may be related to a beam-filling issue, leading to a potential underestimation of the true densities. However, this effect is difficult to calibrate as the whole structure is unresolved. Interferometric observations would reveal the actual angular extent of the structure, allowing for a more accurate estimation of its physical parameters.

\subsection{Origin of the molecular gas}

The morphology and kinematics of the structure raise the question of the origin of the gas. While tori are a quite unusual arrangement for circumstellar material around evolved high mass stars, a few examples are found in the literature, such as the dusty torus around the LBV candidate HD168625 \citep{OHara2003}. There are two possible explanations for the formation of such a structure around a LBV star, namely: (1) the observed molecular gas is stellar ejecta from a process of non-isotropic mass loss. In this scenario the observed CO and \co13 formed when the stellar wind got colder and coagulated; and (2) the molecular material is the relic of the molecular cloud where the star was born, accumulated and compressed by the steady action of a non-spherical wind from a previous stage. 

The total mass of molecular gas present in the CO structure would shed some light on its origin, since very different orders of magnitude are expected in each scenario. This mass can be derived from the column density of H$_2$ via the relation

\begin{equation}
M_\mathrm{mol} = 2d^2m_\mathrm{H}f_\mathrm{He}\int{N(\mathrm{H_{2})}}\mathrm{d}\Omega,
\end{equation}
\label{eq:mass}

\noindent where $d$ is the distance to the object, $m_\mathrm{H}$ the mass of the hydrogen atom, $f_\mathrm{He} = 1.2$ is a correction factor to account for the hydrogen relative abundance, and the last term is the integral of the measured H$_2$ column density over the solid angle subtended by the source. Most of these factors are known or can be estimated with confidence, but the distance and the true H$_2$ column density are subject to a great uncertainty.

\subsubsection{About the distance to MGE 042.0787+00.5084}

There are no available estimates in the literature for the distance to MGE 042.0787+00.5084. By taking the systemic velocity of the structure of 15.4\kms, we calculated two kinematic distances, namely 1.2 and 11.4 kpc. Details on how we determined these distances are provided in Annex \ref{sec:app-a}. We can expect moderate deviations from galactic rotation that would alter these estimates, but even departures of 20\kms{} would increase the near distance just by a factor of $\sim$2, which is not critical for the subsequent interpretation.

Although the two distances are equally possible, we found a number of qualitative arguments that support the near distance: MGE 042.0787+00.5084 is by far the strongest object in the sample studied by \cite{Flagey2014}, being remarkably bright in bands J, H and K, and the symmetry and homogeneity of the nebula suggest a short age, which is more compatible with closer distances.

In order to narrow the range of possible distances, we explored the recently released data from ESA's Gaia mission (Gaia DR2, \citealt{GaiaCol2016}, \citealt{GaiaCol2018}), finding a match within 2\arcsec\ of the source position. This object (SourceID 4307460432455822976) has a parallax of $\pi=0.05\pm0.2$ mas according to Gaia. With such a large uncertainty, distances in the range 4--20 kpc are possible within the error margin. Still, it is possible to deal with inaccurate parallax measurements by means of the Bayesian inference method described by \cite{BailerJones2018}, who applied this statistical approach on the whole DR2 catalog to determine geometric distances according to the following probability density function

\begin{equation}
P(r | \pi, \sigma_\pi, L_\mathrm{sph}) = r^2 \exp\bigg[  -\frac{r}{L_\mathrm{sph}} - \frac{1}{2\sigma_\pi^2} \bigg( \pi - \pi_\mathrm{zp} -\frac{1}{r}\bigg)^2 \bigg],
\end{equation}
\label{eq:bayesian}

\noindent where $\pi$ and $\sigma_\pi$ are the parallax and its uncertainty respectively, $\pi_\mathrm{zp}$ is the parallax zero point and $L_\mathrm{sph}$ is a prior that depends on the galactic coordinates of the source. 

This method has proven to be adequate for narrowing Gaia distances for other LBVs, reaching results in good agreement with literature values even when large uncertainties are involved \citep{Smith2018}. The inferred geometric distance for MGE 042.0787+00.5084 is thus $3.8\substack{+2.2 \\ -1.3}$ kpc, but we must note that this result is based on a very inaccurate parallax and hence must be regarded with caution. 

In the following we adopt a conservative distance of 2.5 kpc, as a compromise between the two approaches and taking into account possible departures from galactic rotation. None the less, we admit that a larger distance is possible (e.g. 10 kpc). The implications of such a long distance on the global parameters of the structure are discussed in Sect. \ref{sec:mass}.

\begin{figure*}
	\includegraphics[width=\textwidth]{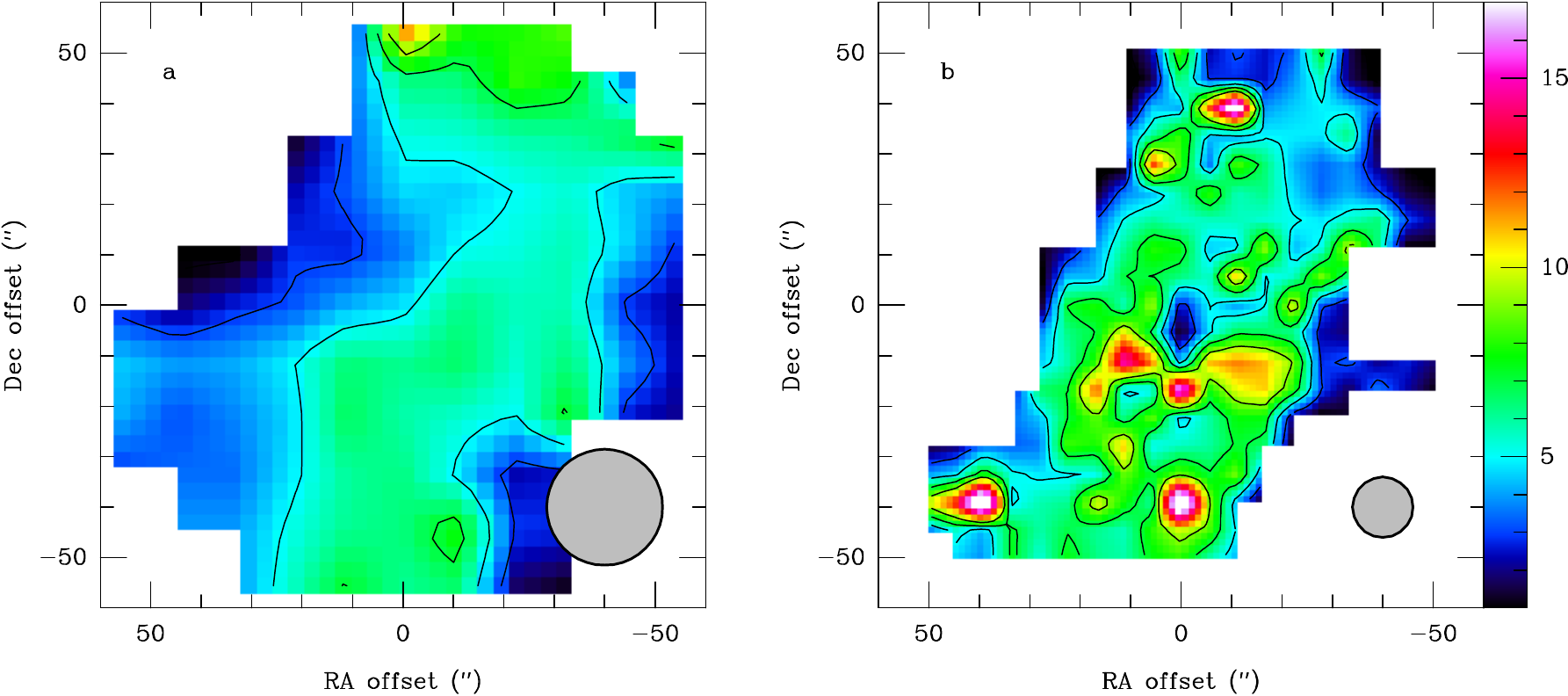}
	\caption{Isotopic ratio maps shown in colour scale: (a) the $^{12}$CO/\co13 $J=1\rightarrow0$ ratio and (b) the $^{12}$CO/\co13  $J=2\rightarrow1$ ratio. The colour scale is the same for the two maps, with contours at 1, 3, 5 and 7. In each map the two lines have been convolved to the same beam size, represented as a gray circle in the bottom right corner. Noisy pixels have been masked out and spatial smoothing has been applied to avoid artefacts.}
	\label{fig:ratios}
\end{figure*}

\subsubsection{Isotopic ratios}
\label{sec:isotopic}

Another essential factor when determining masses is $X$(\co13), the \co13 to H$_2$ relative abundance, which allows to translate $N$(\co13) into $N$(H$_2$). This parameter strongly depends on the $^{13}$C abundance. We observe [$^{12}$CO/\co13] line intensity ratios as low as 5--7 in large regions of the structure, as shown in the line ratio maps of the $1\rightarrow0$ and $2\rightarrow1$ transitions (Fig. \ref{fig:ratios}). According to the LVG results, opacity by itself is not enough to explain an order of magnitude decrease in the [$^{12}$CO/\co13] ratio with respect to the typical ISM values of 70--90 \citep{Wilson1992}. This leaves just two possible explanations for such a low isotopic ratio: a \co12 depletion or a \co13 enhancement.

In the first scenario, chemical fractionation could account for the destruction of \co12. Similarly, a \co13 enhancement could be a direct consequence of stellar evolution. It is known that the CNO cycle favours the production of $^{13}$C in intermediate-mass stars \citep{Berdyugina1994}. The stellar mass range in which this overproduction takes place is not completely constrained, but some studies have measured low [$^{12}$CO/\co13] ratios in a number of massive stars (e.g. $\alpha$ Ori, \citealt{Lambert1984}). If this process occurs in MGE 042.0787+00.5084 as well and the low [$^{12}$C/$^{13}$C] ratio is the result of the central star nucleosynthesis, it would be widespread in the surroundings, therefore indicating that a significant amount of molecular material is of stellar origin.

However, the non-detection of C$^{18}$O in the structure supports the \co13 overproduction hypothesis. We detected this isotopologue in the southeastern cloud, where abundances are close to the canonical values (e.g. [\co13/C$^{18}$O] $\sim$ 5.5, see Fig. \ref{fig:c18o}), but not in the structure, despite having CO and \co13 intensities and rms comparable to those in the cloud. This may imply an excess of \co13. Never the less, other processes could be playing a role: high [\co13/C$^{18}$O] ratios have been proposed as signposts of high FUV fields acting upon the outermost layers of molecular clouds (e.g. photon dominated regions) and causing selective photodissociation of some isotopes \citep{Shimajiri2014}. In this case, this effect would explain the apparent absence of C$^{18}$O, but should only be noticeable locally, in the densest parts of structure, particularly the northwestern clump.

In the $2\rightarrow1$ map (panel b) we note a closed minimum toward the source. The spatial extent of this minimum is comparable to the beam, and the levels of CO and $^{13}$CO in this region are above 5$\sigma$ and 3$\sigma$ respectively. Therefore this minimum is probably a meaningful feature.

Unfortunately, the angular resolution of the 30-m radio telescope does not allow to study this feature in detail and discriminate between the two proposed scenarios. Interferometric observations would be valuable to assess if the low isotopic ratio observed is the result of stellar evolution or a localized effect of the strong UV field of the star. Taking into account the strong uncertainty that affects the $X$(\co13) factor, we use $N$(CO) to estimate the H$_2$ column density, adopting a canonical relative abundance of $X$(CO) = 10$^{-4}$.

\subsubsection{Kinematic model}

Using the global physical parameters as constraints, we developed a simple model to test expanding torus interpretation. The model aimed at reproducing the observed morphology and kinematics following a phenomenological approach. We used the pseudo-Monte Carlo non-LTE radiative transfer code LIME v1.8 \citep{Brinch2010}. LIME simulates photon propagation through unstructured 3D Delaunay grids by applying the SimpleX algorithm \citep{Ritzerveld2006} and assuming ballistic transport. For a given molecule and transition LIME calculates level populations and generates synthetic spectral cubes, based on the LAMDA collisional rates. We processed these cubes with a custom Python pipeline to simulate the observation with the IRAM 30-m telescope, obtaining maps directly comparable to the original data. We used the CO $J=2\rightarrow1$ line because it provides the highest angular resolution and signal-to-noise ratio.

We modelled the structure as an expanding torus characterized by a set of geometrical and physical parameters. A complete list of the parameters is provided in Table \ref{tab:lime-params} and the model is described in detail in Annex \ref{sec:app-b}. Geometry is controlled by distance, inclination and position angle. We defined $r_0$ as a scaling radius coincident with the inner edge of the torus to parametrize density and temperature. Density is described by a toroidal function where the parameter $\theta_H$, the half-opening angle, determines the oblateness of the structure and the dimensionless parameters $p$ and $q$ dominate the steepness of the density gradient in the radial and angular directions. Since the torus is unresolved in the 30-m data we do not know its actual thickness, so we set some of these parameters manually for convenience. For the temperature profile we used a segmented function, which remains constant at a reference temperature in the cavity and behaves as a radial power law for $r > r_0$. Regarding the velocity field, we considered a purely radial expansion. 
\subsubsection{Model results} \label{sec:modelresults}

We adopted a distance of 2.5 kpc with a relative abundance $X$(CO) = $10^{-4}$ as explained above. The free parameters of the model are the reference density $\rho_0$ --related to $n(\mathrm{H_2})$--, the reference temperature $T_0$ --related to the kinetic temperature of the gas--, the expansion velocity $V_\mathrm{exp}$ and the inclination angle $i$. To account for the uncertainty in the determination of the inclination angle and the possibility of a nearly face-on structure discussed in Sect. \ref{sec:discussion:morfkin}, we built three types of models, A, B and C, corresponding to inclinations of 15, 30 and 45\degr\ respectively. We used reference temperatures in the range 10--70 K and densities of H$_2$ in the range 0.8--4$\times10^3$ cm$^{-3}$, in agreement with the LVG results (see Table \ref{tab:cd-radex}). We also let the expansion velocity vary from 1 to 5 \kms\ to cover possible scenarios where a low inclination somehow compensates for a higher velocity. 

\begin{figure*}
	\includegraphics[width=\textwidth]{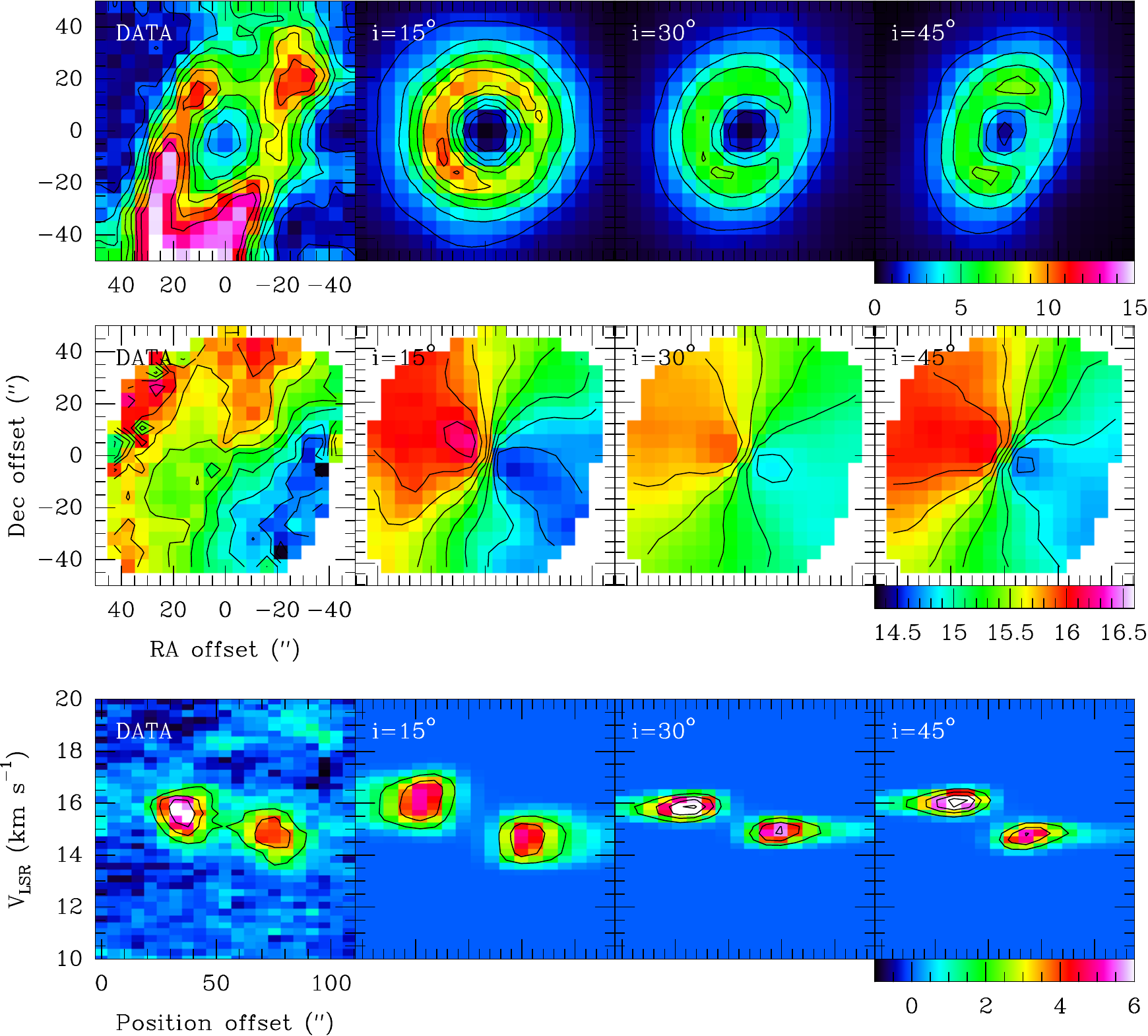}
	\caption{Comparison between data and best model for each inclination angle (see Table \ref{tab:best-models}). Top row: CO $J=2\rightarrow1$ integrated line intensity maps in the velocity range (+13.4, +17.4) \kms. Colour scale in units of K \kms, with contours starting at 5$\sigma$ with steps of $5\sigma$. Middle row: Intensity-weighted velocity maps in the same velocity range. Bottom row: Position-velocity diagrams in the velocity range (+10, +20) \kms\ for a slice with a P.A. of 60\degr (see Fig. \ref{fig:intensity}).}
	\label{fig:model}
\end{figure*}

We explored the parameter space iteratively looking for the best fit in the velocity range (+13.4, +17.4)\kms, based on a channel-wise $\chi^2$ criteria described in Annex \ref{sec:app-b}. The best-fitting set of parameters for each kind of model is presented in Table \ref{tab:best-models} and a comparison of the main observables (line intensity maps, velocity fields and position-velocity diagrams) is shown in Fig. \ref{fig:model}. The three models show a preference towards warm, low density solutions, as they reach the best fit for $\rho_0 = 0.8\times10^{3}$ cm$^{-3}$ and $T_0$ = 70 K. It is important to note that these are reference values, so the average $T_\mathrm{k}$ and $n(\mathrm{H_2})$ in the structure may be slightly lower. In any case, the best-fitting density is not particularly high, which could explain the nondetection of other less abundant molecules in the structure. Other denser and colder models fail to match the observed intensities, even by an order of magnitude.

Model A, which corresponds to an inclination of 15\degr, provides the best match, with a quite convincing reproduction of the morphology and kinematics of the gas. The expansion velocity of 3 \kms\ is $\sim$2 times higher that the prior estimates from the PV diagrams in Fig. \ref{fig:pv-cuts}, but still much lower than typical values. The integrated intensity map is slightly weaker than the data in most of the structure, but this difference is an expected side effect of the simpleness of our torus model and the uniformity of the density distribution. Never the less, model A is able to reproduce with high fidelity the observed velocity gradient everywhere but in the cavity, which in our model is highly depleted. Models B and C, corresponding to higher inclinations, fail to reproduce the observed morphology as the resulting structure is exceedingly elongated. Similarly, while the velocity gradient approaches that of model A, the velocity extent of the emission blobs in the PV diagrams is smaller than expected, of just 0.75--1 \kms, which means that the material in the line of sight is confined in a narrower velocity range.

\subsection{Mass and mass-loss} \label{sec:mass}

In view of these results, we conclude that the structure of molecular gas seen around MGE 042.0787+00.5084 is a low-density, nearly face-on torus that expands slowly. Adopting a conservative $T_\mathrm{k}$ of 50 K, we derive a total mass for the structure of $0.6 \pm 0.1$ M$_{\sun}$ ($0.5 \pm 0.1$ M$_{\sun}$ in the torus and 0.1 M$_{\sun}$ in the clump). This mass is probably rather low to be the remnant of a molecular cloud piled up by the stellar wind, but it is compatible with masses of molecular gas found around other LBV stars \citep{Rizzo2008}. 

Therefore, assuming all the gas formed from stellar ejecta, a crude estimate of the mass loss rate of MGE 042.0787+00.5084 can be provided simply by calculating the dynamical age of the circumstellar structure. The torus deprojected radius is $\sim$ 15\arcsec\ on average, which at 2.5 kpc translates into 0.18 pc. For $V_\mathrm{exp} =$ 3 \kms, the dynamical age of the structure is

\begin{equation}
t_\mathrm{dyn} < \frac{R_\mathrm{inner}}{V_\mathrm{exp}} \sim  6\times10^4 \mathrm{yrs.}
\end{equation}
\label{eq:tdyn}

The LBV phase typically lasts for a few $10^4$ years, so this dynamical age is compatible with a structure formed during this period. However, the age derived by this method must be regarded as a coarse upper limit, as the expansion velocity of the gas could have been significantly higher in the past, implying a younger structure.

Never the less this time-scale is useful to estimate the average mass-loss rate of the star during the LBV phase, yielding a value of $\dot{M} =$ 0.8--1.2$\times10^{-5}$ \myr.  This result is compatible with evolutionary models \citep{Langer1994}, laying in the lower end of the LBV spectrum, where stars exhibit high temperatures and moderate mass-loss rates \citep{Lamers1989}. 

Still, the mass-loss rate estimate provided here must be interpreted carefully. It is a time average over a period of $6\times10^4$ years, so the star could have experienced brief periods of severe mass-loss in the past and we cannot use it to infer its current state. In addition, we only considered molecular gas in the calculation, which most likely is not the only contributor to the total mass. Other significant components (e.g. neutral and ionized gas) would increase the mass-loss rate and were not taken into account.

Finally, it is worth to mention that the derived time-scale for the molecular gas structure also constrains the age of the inner nebula. While the torus is the result of steady stellar winds, the infrared nebula may trace a much more recent eruptive event. The possibility of simultaneously observing two successive mass-loss events traced by different phenomena makes MGE 042.0787+00.5084 a very interesting target for further studies.

In our analysis we have adopted a conservative estimate of 2.5 kpc based on a series of reasonable assumptions from the scarce data available. We find that larger distances, in the range of 10 kpc or more, are still possible --according to Gaia parallaxes and admitting feasible departures from galactic rotation-- but somehow less likely, as they lead to scenarios difficult to reconcile with the observations.

Size and age of the structure increase with $d$ while mass scales with $d^2$. Therefore, at 10 kpc the structure would be $2.4\times10^5$ years old for $V_\mathrm{exp} = 3$ \kms, which places its origin into the main sequence stage. The total mass of molecular gas in the torus would then be $9.6 \pm 1.6$ M$_{\sun}$.  \cite{Groh2014} analyzed the evolution of a non-rotating 60 M$_{\sun}$ star in terms of mass-loss, finding that the mass lost during the 3 Myr prior to the LBV phase is less than 10 M$_{\sun}$ in total. Considering that molecular material represents just a fraction of the mass ejected, MGE 042.0787+00.5084 would need to rival the luminosity of $\eta$ Car and should be an extremely efficient molecule producer in order to form such a massive structure. In addition, with the contribution of the inner nebula, the circumstellar material around this star would probably exceed the mass of any other known LBV nebula. On the other hand, the density of the gas is proportional to $d^{-1}$; as a consequence, at 10 kpc the gas would be diffuse, with an average $n(\mathrm{H}_2)$ of $\sim80$ cm$^{-3}$. Under these conditions, LTE modelling would require extremely high kinetic temperatures far from the star (above 400 K at 0.75 pc) to reproduce the observed CO line intensities.

These are only indirect arguments that do not allow us to rule out far distances but stress the need for an accurate distance determination, which would permit to trace back the evolution and mass-loss record of this source.

\begin{table}
\centering
\caption{LIME model parametrization. Free parameters are expressed as a range.}
\label{tab:lime-params}
\begin{tabular}{llll}
\hline
\textbf{Parameter} & \textbf{Meaning}& \textbf{Value}& \textbf{Units}	\\
\hline
d 					& distance & 2.5 & kpc	\\
$r_0$      			& inner radius & 0.18 & pc	\\
p, q					& density steepness & 2 &	\\ 
$\gamma$   			& temperature steepness &1.5	& \\
$\theta_\mathrm{H}$	& half-opening angle & 15	& deg	\\ 
\hline
i  					& inclination& 15--45	& deg	\\
$T_0$  				& temperature at $r_0$	& 10--80		& K	\\
$\rho_0$ 			&density at $r_0$	& 0.8--4$\times10^3$	& cm$^{-3}$  	 \\
$V_\mathrm{exp}$		&expansion velocity & 1--5	&\kms \\ \hline
\end{tabular}
\end{table}

\begin{table}
\centering
\caption{Best-fitting parameters for each LIME model.}
\label{tab:best-models}
\begin{tabular}{lccc}
\hline
& \multicolumn{3}{c}{\textbf{Model}} \\
\textbf{Parameter} &A (i = 15\degr) & B (i = 30\degr) & C (i = 45\degr) \\
\hline
$\rho_0$ [cm$^{-3}$]& 800 & 800 & 800 \\
$T_0 $ [K] & 70 & 70 & 70 \\
$V_\mathrm{exp}$ [\kms] & 3 & 1 & 1 \\
\hline
\end{tabular}
\end{table}

\section{Conclusions}\label{sec:conclusions}

We report the first detection of CO gas associated with the candidate LBV MGE 042.0787+00.5084 forming a circumstellar structure that encloses the infrared nebula. This detection was obtained by means of 1.3 and 3 mm observations in a region of $1\farcm5\times1\farcm5$ in the field of the source carried out with the IRAM 30-m telescope. After a detailed study of the morphology and kinematics of the structure, we estimate the global physical parameters of the gas by LTE and non-LTE modelling. Finally we provide an interpretation for the formation of the molecular gas consistent with two scenarios in the LBV phase. The key findings in this paper are:

\begin{enumerate}
\item We found significant emission of CO and \co13 $J=1\rightarrow0$ and $J=2\rightarrow1$ in the velocity range (+13.4, +17.4)\kms, tracing a structure with a clear circumstellar distribution. We interpreted this structure as an expanding torus according to the observed morphology and kinematics.
\item We studied the physical conditions in two separate regions of the structure, namely the northwestern clump and the main torus. We constrained density, column density and kinetic temperature under LTE and non-LTE assumptions, finding that the clump is denser than the main torus by a factor of 1.5--2 regardless of the temperature, with H$_2$ volume densities of a few $10^3$ cm$^{-3}$. On the other hand, only kinetic temperatures in the range 10--70 K  for the torus and 10--100 K for the clump are consistent with the observed line intensities and ratios.
\item The measured [$^{12}$CO/\co13] ratio was rather low, in the range from 5 to 7 in most of the structure. This result was particularly intriguing as no C$^{18}$O was detected, except in the southern cloud. We proposed a \co13 overproduction as an explanation for such an abnormal isotopic ratio, discussing the implications of this enhancement as the result of stellar evolution or selective photodissociation. 
\item The formation of a toroidal structure around a LBV star can be explained by two scenarios: (1) molecular ejecta from the star via non-spherical mass-loss, or (2) compression of the parent molecular cloud due to the effects of the strong and steady stellar wind. The determination of the total amount of molecular material could pinpoint the most likely scenario, but the distance to the source is poorly constrained. By adopting $d$ = 2.5 kpc, we derived a mass of molecular gas of 0.6 M$_{\sun}$, consistent with an average mass-loss rate of 0.8--1.2 $\times10^{-5}$\myr during the LBV phase.
\item We developed a simple kinematic model with the aim of reproducing the morphology and velocity fields of the structure. We studied the influence of density, temperature, inclination and expansion velocity. The model with the lowest inclination ($i$=15\degr) and a rather low expansion velocity (3\kms) provided the best fit, being able to accurately reproduce the observed velocity gradient in most of the structure, as proven by first-order moments and PV diagrams. This supports the interpretation of the structure as an expanding torus of CO with a nearly face-on viewing geometry. 
\end{enumerate}
 
The characterization of the molecular material associated with MGE 042.0787+00.5084 constitutes an excellent example of how studies of this kind provide useful information on the mass-loss processes during the LBV phase. This source has shown a great potential to learn about the interplay between evolved high mass stars and the ISM, deserving follow-up studies to better constrain its physical parameters and evolutionary record. This can be achieved by means of interferometric observations and near-infrared high-resolution spectroscopy. Most importantly, this successful detection proves that the existence of molecular material in the vicinity of evolved massive stars is not an exceptional finding but likely a common feature, therefore laying the foundations for future studies involving similar sources and less abundant molecules.

\section*{Acknowledgements}

This work is based on observations carried out under project number 043-17 with the IRAM 30-m telescope. IRAM is supported by INSU/CNRS (France), MPG (Germany) and IGN (Spain). We acknowledge the IRAM staff from Pico Veleta for the help provided during the observations. J.R.R. acknowledges the support from project ESP2015-65597-C4-1-R (MINECO/FEDER).



\bibliographystyle{mnras}
\bibliography{biblio} 

\begin{thebibliography}{}
\makeatletter
\relax
\def\mn@urlcharsother{\let\do\@makeother \do\$\do\&\do\#\do\^\do\_\do\%\do\~}
\def\mn@doi{\begingroup\mn@urlcharsother \@ifnextchar [ {\mn@doi@}
  {\mn@doi@[]}}
\def\mn@doi@[#1]#2{\def\@tempa{#1}\ifx\@tempa\@empty \href
  {http://dx.doi.org/#2} {doi:#2}\else \href {http://dx.doi.org/#2} {#1}\fi
  \endgroup}
\def\mn@eprint#1#2{\mn@eprint@#1:#2::\@nil}
\def\mn@eprint@arXiv#1{\href {http://arxiv.org/abs/#1} {{\tt arXiv:#1}}}
\def\mn@eprint@dblp#1{\href {http://dblp.uni-trier.de/rec/bibtex/#1.xml}
  {dblp:#1}}
\def\mn@eprint@#1:#2:#3:#4\@nil{\def\@tempa {#1}\def\@tempb {#2}\def\@tempc
  {#3}\ifx \@tempc \@empty \let \@tempc \@tempb \let \@tempb \@tempa \fi \ifx
  \@tempb \@empty \def\@tempb {arXiv}\fi \@ifundefined
  {mn@eprint@\@tempb}{\@tempb:\@tempc}{\expandafter \expandafter \csname
  mn@eprint@\@tempb\endcsname \expandafter{\@tempc}}}

\bibitem[\protect\citeauthoryear{{Bailer-Jones}, {Rybizki}, {Fouesneau},
  {Mantelet}  \& {Andrae}}{{Bailer-Jones} et~al.}{2018}]{BailerJones2018}
{Bailer-Jones} C.~A.~L.,  {Rybizki} J.,  {Fouesneau} M.,  {Mantelet} G.,
  {Andrae} R.,  2018, preprint, \href
  {http://adsabs.harvard.edu/abs/2018arXiv180410121B} {} (\mn@eprint {arXiv}
  {1804.10121})

\bibitem[\protect\citeauthoryear{{Berdyugina} \& {Savanov}}{{Berdyugina} \&
  {Savanov}}{1994}]{Berdyugina1994}
{Berdyugina} S.~V.,  {Savanov} I.~S.,  1994, Astronomy Letters, \href
  {http://adsabs.harvard.edu/abs/1994AstL...20..639B} {20, 639}

\bibitem[\protect\citeauthoryear{{Brinch} \& {Hogerheijde}}{{Brinch} \&
  {Hogerheijde}}{2010}]{Brinch2010}
{Brinch} C.,  {Hogerheijde} M.~R.,  2010, \mn@doi [\aap]
  {10.1051/0004-6361/201015333}, \href
  {http://adsabs.harvard.edu/abs/2010A%26A...523A..25B} {523, A25}

\bibitem[\protect\citeauthoryear{{Buemi} et~al.,}{{Buemi}
  et~al.}{2017}]{Buemi2017}
{Buemi} C.~S.,  et~al., 2017, \mn@doi [\mnras] {10.1093/mnras/stw3074}, \href
  {http://adsabs.harvard.edu/abs/2017MNRAS.465.4147B} {465, 4147}

\bibitem[\protect\citeauthoryear{{Clark}, {Larionov}  \& {Arkharov}}{{Clark}
  et~al.}{2005}]{Clark2005}
{Clark} J.~S.,  {Larionov} V.~M.,   {Arkharov} A.,  2005, \mn@doi [\aap]
  {10.1051/0004-6361:20042563}, \href
  {http://adsabs.harvard.edu/abs/2005A%26A...435..239C} {435, 239}

\bibitem[\protect\citeauthoryear{{Davidson}}{{Davidson}}{1989}]{Davidson1989}
{Davidson} K.,  1989, in {Davidson} K.,  {Moffat} A.~F.~J.,   {Lamers}
  H.~J.~G.~L.~M.,  eds,  Astrophysics and Space Science Library Vol. 157, IAU
  Colloq. 113: Physics of Luminous Blue Variables. pp 101--107,
  \mn@doi{10.1007/978-94-009-1031-7_11}

\bibitem[\protect\citeauthoryear{{Fich}, {Blitz}  \& {Stark}}{{Fich}
  et~al.}{1989}]{Fich1989}
{Fich} M.,  {Blitz} L.,   {Stark} A.~A.,  1989, \mn@doi [\apj]
  {10.1086/167591}, \href {http://adsabs.harvard.edu/abs/1989ApJ...342..272F}
  {342, 272}

\bibitem[\protect\citeauthoryear{{Flagey}, {Noriega-Crespo}, {Petric}  \&
  {Geballe}}{{Flagey} et~al.}{2014}]{Flagey2014}
{Flagey} N.,  {Noriega-Crespo} A.,  {Petric} A.,   {Geballe} T.~R.,  2014,
  \mn@doi [\aj] {10.1088/0004-6256/148/2/34}, \href
  {http://adsabs.harvard.edu/abs/2014AJ....148...34F} {148, 34}

\bibitem[\protect\citeauthoryear{{Gaia Collaboration} et~al.,}{{Gaia
  Collaboration} et~al.}{2016}]{GaiaCol2016}
{Gaia Collaboration} et~al., 2016, \mn@doi [\aap]
  {10.1051/0004-6361/201629272}, \href
  {http://adsabs.harvard.edu/abs/2016A%26A...595A...1G} {595, A1}

\bibitem[\protect\citeauthoryear{{Gaia Collaboration}, {Brown}, {Vallenari},
  {Prusti}, {de Bruijne}, {Babusiaux}  \& {Bailer-Jones}}{{Gaia Collaboration}
  et~al.}{2018}]{GaiaCol2018}
{Gaia Collaboration} {Brown} A.~G.~A.,  {Vallenari} A.,  {Prusti} T.,  {de
  Bruijne} J.~H.~J.,  {Babusiaux} C.,   {Bailer-Jones} C.~A.~L.,  2018,
  preprint, \href {http://adsabs.harvard.edu/abs/2018arXiv180409365G} {}
  (\mn@eprint {arXiv} {1804.09365})

\bibitem[\protect\citeauthoryear{{Groh}, {Meynet}  \& {Ekstr{\"o}m}}{{Groh}
  et~al.}{2013}]{Groh2013}
{Groh} J.~H.,  {Meynet} G.,   {Ekstr{\"o}m} S.,  2013, \mn@doi [\aap]
  {10.1051/0004-6361/201220741}, \href
  {http://adsabs.harvard.edu/abs/2013A%26A...550L...7G} {550, L7}

\bibitem[\protect\citeauthoryear{{Groh}, {Meynet}, {Ekstr{\"o}m}  \&
  {Georgy}}{{Groh} et~al.}{2014}]{Groh2014}
{Groh} J.~H.,  {Meynet} G.,  {Ekstr{\"o}m} S.,   {Georgy} C.,  2014, \mn@doi
  [\aap] {10.1051/0004-6361/201322573}, \href
  {http://adsabs.harvard.edu/abs/2014A%26A...564A..30G} {564, A30}

\bibitem[\protect\citeauthoryear{{Guilloteau} \& {Lucas}}{{Guilloteau} \&
  {Lucas}}{2000}]{Guilloteau2000}
{Guilloteau} S.,  {Lucas} R.,  2000, in {Mangum} J.~G.,  {Radford} S.~J.~E.,
  eds,  Astronomical Society of the Pacific Conference Series Vol. 217, Imaging
  at Radio through Submillimeter Wavelengths. p.~299

\bibitem[\protect\citeauthoryear{{Gvaramadze} et~al.,}{{Gvaramadze}
  et~al.}{2012}]{Gvaramadze2012}
{Gvaramadze} V.~V.,  et~al., 2012, \mn@doi [\mnras]
  {10.1111/j.1365-2966.2012.20556.x}, \href
  {http://adsabs.harvard.edu/abs/2012MNRAS.421.3325G} {421, 3325}

\bibitem[\protect\citeauthoryear{{Humphreys} \& {Davidson}}{{Humphreys} \&
  {Davidson}}{1994}]{Humphreys1994}
{Humphreys} R.~M.,  {Davidson} K.,  1994, \mn@doi [\pasp] {10.1086/133478},
  \href {http://adsabs.harvard.edu/abs/1994PASP..106.1025H} {106, 1025}

\bibitem[\protect\citeauthoryear{{Ingallinera} et~al.,}{{Ingallinera}
  et~al.}{2014}]{Ingallinera2014}
{Ingallinera} A.,  et~al., 2014, \mn@doi [\mnras] {10.1093/mnras/stt2157},
  \href {http://adsabs.harvard.edu/abs/2014MNRAS.437.3626I} {437, 3626}

\bibitem[\protect\citeauthoryear{{Ingallinera} et~al.,}{{Ingallinera}
  et~al.}{2016}]{Ingallinera2016}
{Ingallinera} A.,  et~al., 2016, \mn@doi [\mnras] {10.1093/mnras/stw2053},
  \href {http://adsabs.harvard.edu/abs/2016MNRAS.463..723I} {463, 723}

\bibitem[\protect\citeauthoryear{{Lambert}, {Brown}, {Hinkle}  \&
  {Johnson}}{{Lambert} et~al.}{1984}]{Lambert1984}
{Lambert} D.~L.,  {Brown} J.~A.,  {Hinkle} K.~H.,   {Johnson} H.~R.,  1984,
  \mn@doi [\apj] {10.1086/162401}, \href
  {http://adsabs.harvard.edu/abs/1984ApJ...284..223L} {284, 223}

\bibitem[\protect\citeauthoryear{{Lamers}}{{Lamers}}{1989}]{Lamers1989}
{Lamers} H.~J.~G.~L.~M.,  1989, in {Davidson} K.,  {Moffat} A.~F.~J.,
  {Lamers} H.~J.~G.~L.~M.,  eds,  Astrophysics and Space Science Library Vol.
  157, IAU Colloq. 113: Physics of Luminous Blue Variables. pp 135--146,
  \mn@doi{10.1007/978-94-009-1031-7_17}

\bibitem[\protect\citeauthoryear{{Langer}, {Hamann}, {Lennon}, {Najarro},
  {Pauldrach}  \& {Puls}}{{Langer} et~al.}{1994}]{Langer1994}
{Langer} N.,  {Hamann} W.-R.,  {Lennon} M.,  {Najarro} F.,  {Pauldrach}
  A.~W.~A.,   {Puls} J.,  1994, \aap, \href
  {http://adsabs.harvard.edu/abs/1994A%26A...290..819L} {290, 819}

\bibitem[\protect\citeauthoryear{{Marston} \& {McCollum}}{{Marston} \&
  {McCollum}}{2006}]{Marston2006}
{Marston} A.~P.,  {McCollum} B.,  2006, in {Kraus} M.,  {Miroshnichenko} A.~S.,
   eds,  Astronomical Society of the Pacific Conference Series Vol. 355, Stars
  with the B[e] Phenomenon. p.~189

\bibitem[\protect\citeauthoryear{{Mizuno} et~al.,}{{Mizuno}
  et~al.}{2010}]{Mizuno2010}
{Mizuno} D.~R.,  et~al., 2010, \mn@doi [\aj] {10.1088/0004-6256/139/4/1542},
  \href {http://adsabs.harvard.edu/abs/2010AJ....139.1542M} {139, 1542}

\bibitem[\protect\citeauthoryear{{O'Hara}, {Meixner}, {Speck}, {Ueta}  \&
  {Bobrowsky}}{{O'Hara} et~al.}{2003}]{OHara2003}
{O'Hara} T.~B.,  {Meixner} M.,  {Speck} A.~K.,  {Ueta} T.,   {Bobrowsky} M.,
  2003, \mn@doi [\apj] {10.1086/379058}, \href
  {http://adsabs.harvard.edu/abs/2003ApJ...598.1255O} {598, 1255}

\bibitem[\protect\citeauthoryear{{Ossenkopf} \& {Henning}}{{Ossenkopf} \&
  {Henning}}{1994}]{Ossenkopf1994}
{Ossenkopf} V.,  {Henning} T.,  1994, \aap, \href
  {http://adsabs.harvard.edu/abs/1994A%26A...291..943O} {291, 943}

\bibitem[\protect\citeauthoryear{Ritzerveld \& Icke}{Ritzerveld \&
  Icke}{2006}]{Ritzerveld2006}
Ritzerveld J.,  Icke V.,  2006, \mn@doi [Phys. Rev. E]
  {10.1103/PhysRevE.74.026704}, 74, 026704

\bibitem[\protect\citeauthoryear{{Rizzo}, {Jim{\'e}nez-Esteban}  \&
  {Ortiz}}{{Rizzo} et~al.}{2008}]{Rizzo2008}
{Rizzo} J.~R.,  {Jim{\'e}nez-Esteban} F.~M.,   {Ortiz} E.,  2008, \mn@doi
  [\apj] {10.1086/588455}, \href
  {http://adsabs.harvard.edu/abs/2008ApJ...681..355R} {681, 355}

\bibitem[\protect\citeauthoryear{{Rizzo}, {Palau}, {Jim{\'e}nez-Esteban}  \&
  {Henkel}}{{Rizzo} et~al.}{2014}]{Rizzo2014}
{Rizzo} J.~R.,  {Palau} A.,  {Jim{\'e}nez-Esteban} F.,   {Henkel} C.,  2014,
  \mn@doi [\aap] {10.1051/0004-6361/201323170}, \href
  {http://adsabs.harvard.edu/abs/2014A%26A...564A..21R} {564, A21}

\bibitem[\protect\citeauthoryear{{S{\'a}nchez Contreras} et~al.,}{{S{\'a}nchez
  Contreras} et~al.}{2015}]{SanchezContreras2015}
{S{\'a}nchez Contreras} C.,  et~al., 2015, \mn@doi [\aap]
  {10.1051/0004-6361/201525652}, \href
  {http://adsabs.harvard.edu/abs/2015A%26A...577A..52S} {577, A52}

\bibitem[\protect\citeauthoryear{{Sch{\"o}ier}, {van der Tak}, {van Dishoeck}
  \& {Black}}{{Sch{\"o}ier} et~al.}{2005}]{Schoier2005}
{Sch{\"o}ier} F.~L.,  {van der Tak} F.~F.~S.,  {van Dishoeck} E.~F.,   {Black}
  J.~H.,  2005, \mn@doi [\aap] {10.1051/0004-6361:20041729}, \href
  {http://esoads.eso.org/abs/2005A%26A...432..369S} {432, 369}

\bibitem[\protect\citeauthoryear{{Shimajiri} et~al.,}{{Shimajiri}
  et~al.}{2014}]{Shimajiri2014}
{Shimajiri} Y.,  et~al., 2014, \mn@doi [\aap] {10.1051/0004-6361/201322912},
  \href {http://adsabs.harvard.edu/abs/2014A%26A...564A..68S} {564, A68}

\bibitem[\protect\citeauthoryear{{Smith}, {Aghakhanloo}, {Murphy}, {Stassun},
  {Drout}  \& {Groh}}{{Smith} et~al.}{2018}]{Smith2018}
{Smith} N.,  {Aghakhanloo} M.,  {Murphy} J.~W.,  {Stassun} K.~G.,  {Drout}
  M.~R.,   {Groh} J.~H.,  2018, preprint, \href
  {http://adsabs.harvard.edu/abs/2018arXiv180503298S} {} (\mn@eprint {arXiv}
  {1805.03298})

\bibitem[\protect\citeauthoryear{{Sobolev}}{{Sobolev}}{1960}]{Sobolev1960}
{Sobolev} V.~V.,  1960, {Moving envelopes of stars}

\bibitem[\protect\citeauthoryear{{Stringfellow}, {Gvaramadze}, {Beletsky}  \&
  {Kniazev}}{{Stringfellow} et~al.}{2012}]{Stringfellow2012}
{Stringfellow} G.~S.,  {Gvaramadze} V.~V.,  {Beletsky} Y.,   {Kniazev} A.~Y.,
  2012, in {Richards} M.~T.,  {Hubeny} I.,  eds,  IAU Symposium Vol. 282, From
  Interacting Binaries to Exoplanets: Essential Modeling Tools. pp 267--268
  (\mn@eprint {arXiv} {1112.2685}), \mn@doi{10.1017/S1743921311027542}

\bibitem[\protect\citeauthoryear{{Umana}, {Buemi}, {Trigilio}, {Leto}, {Hora}
  \& {Fazio}}{{Umana} et~al.}{2011}]{Umana2011}
{Umana} G.,  {Buemi} C.~S.,  {Trigilio} C.,  {Leto} P.,  {Hora} J.~L.,
  {Fazio} G.,  2011, Bulletin de la Societe Royale des Sciences de Liege, \href
  {http://adsabs.harvard.edu/abs/2011BSRSL..80..335U} {80, 335}

\bibitem[\protect\citeauthoryear{{Wilson}}{{Wilson}}{2009}]{Wilson2009}
{Wilson} T.~L.,  2009, preprint, \href
  {http://adsabs.harvard.edu/abs/2009arXiv0903.0562W} {} (\mn@eprint {arXiv}
  {0903.0562})

\bibitem[\protect\citeauthoryear{{Wilson} \& {Matteucci}}{{Wilson} \&
  {Matteucci}}{1992}]{Wilson1992}
{Wilson} T.~L.,  {Matteucci} F.,  1992, \mn@doi [\aapr] {10.1007/BF00873568},
  \href {http://adsabs.harvard.edu/abs/1992A%26ARv...4....1W} {4, 1}

\bibitem[\protect\citeauthoryear{{van Boekel} et~al.,}{{van Boekel}
  et~al.}{2003}]{VanBoekel2003}
{van Boekel} R.,  et~al., 2003, \mn@doi [\aap] {10.1051/0004-6361:20031500},
  \href {http://adsabs.harvard.edu/abs/2003A%26A...410L..37V} {410, L37}

\bibitem[\protect\citeauthoryear{{van der Tak}, {Black}, {Sch{\"o}ier},
  {Jansen}  \& {van Dishoeck}}{{van der Tak} et~al.}{2007}]{Tak2007}
{van der Tak} F.~F.~S.,  {Black} J.~H.,  {Sch{\"o}ier} F.~L.,  {Jansen} D.~J.,
   {van Dishoeck} E.~F.,  2007, \mn@doi [\aap] {10.1051/0004-6361:20066820},
  \href {http://adsabs.harvard.edu/abs/2007A%26A...468..627V} {468, 627}

\makeatother
\end{thebibliography}



\appendix

\section{Computation of kinematic distances}
\label{sec:app-a}

Assuming MGE 042.0787+00.5084 is in circular rotation around the Galactic Centre, its angular velocity is given by 

\begin{equation}
\omega = \frac{V_\mathrm{LSR}}{R_0 \sin{l} \cos{b}}+\omega_0
\label{eq:galw}
\end{equation}

\noindent where $l$ and $b$ denote the galactic longitude and latitude of the source and $R_0$ and $\omega_0$ are the galactocentric distance and angular velocity of the Sun respectively. Considering the fit for the rotation curve of the Milky Way provided by \cite{Fich1989}, the galactocentric radius of an object with angular velocity $\omega$ is

\begin{equation}
R = \frac{1.00746R_0}{\frac{\omega}{\omega_0}+0.017112}
\label{eq:galr}
\end{equation}

This galactocentric radius $R$ relates to the distance through the expression

 \begin{equation}
R = (R_0^2+d^2-2R_0d\cos{l})^{1/2}
\label{eq:gald}
\end{equation}

\noindent which is a regular second degree equation and thus admits two possible solutions. Taking the canonical values for the local galactic constants $R_0=8.5$ kpc and $\Theta_0=220$ \kms, we obtain a galactocentric radius of R = 7.63 kpc for MGE 042.0787+00.5084, which corresponds with $d_\mathrm{near}= 1.2$ kpc and $d_\mathrm{far} = 11.4$ kpc.

\section{LIME model}
\label{sec:app-b}

The model is an ideal, axisymmetric torus defined in a cylindrical coordinate system $(r, \phi, z)$. Geometry of the model is defined by three parameters: distance, position angle and inclination angle of the structure with respect to the line of sight. 

We used ortho- and para- H$_2$ as the main collision partners for photon propagation, with an ortho-para ratio of 3, and grains covered by thin ice mantles for dust opacity \citep{Ossenkopf1994}. A CO relative abundance of $X_\mathrm{CO} = 10^{-4}$ was used. No magnetic field was included in the model. 

The model assumes gas and dust are thermally coupled, following the same torus-like density distribution, such that

\begin{equation}
\rho(r,z)= \begin{cases}
\rho_0 \left( \frac{r}{r_0}\right)^{-p} \exp\bigg[ -q |\cos\theta| \bigg]  &\quad\text{if $r \geq r_0$, $\theta \leq \theta_\mathrm{H}$}\\
\delta\rho_0 &\quad\text{otherwise}
\end{cases}
\label{eq:mod-dens}
\end{equation}

\noindent where $r_0$ is a scaling radius equal to the inner radius of the torus, $\rho_0$ is a reference density value, $p$ and $q$ are dimensionless parameters that determine the radial and angular density gradients respectively, and $\theta_\mathrm{H}$ is the half-opening angle of the torus. To model the inner cavity we added a depletion factor $\delta = 10^{-5}$ that weights the reference density when $r \leq r_0$. Accordingly, points with $\theta \geq \theta_\mathrm{H}$ (i.e. outside the torus) are similarly depleted.

We modelled the temperature profile with a simple radial power law controlled by the dimensionless parameter $\gamma$,

\begin{equation}
T(r) = T_0 \left(\frac{r}{r_0}\right)^{\gamma}
\label{eq:mod-temp}
\end{equation}

\noindent for $r \geq r_0$, remaining constant at reference temperature $T_0$ within the cavity, which is consistent with the absence of material to absorb radiation. 

Regarding the torus kinematics, we adopted a velocity field consistent with a structure expanding at a constant rate, as discussed in Sect. \ref{sec:discussion:morfkin}. Consequently the velocity vector is completely radial in the model space, with $|\vec{v}| = V_\mathrm{exp}$.

To measure the goodness of the fitting we used a channel-wise $\chi^2$ criteria in the velocity range of interest (+13.4, +17.4 \kms), defined as 

\begin{equation}
\chi^2 = \sum_{j}\frac{(T^{\mathrm{model}}_j-T^{\mathrm{data}}_j)^2}{\sigma^2}
\end{equation}

\noindent where $T^{\mathrm{model}}$ and $T^{\mathrm{data}}$ are the main-beam temperature per pixel for model and data respectively, $\sigma$ is the data rms scaling factor and $j$ denotes the velocity channel. By doing so we account for the morphology and velocity extent of the emission simultaneously.


\bsp	
\label{lastpage}
\end{document}